\documentclass[twocolumn]{assets/apie}
\pdfoutput=1
\usepackage{listings}
\makeatletter
\newcommand{\removelatexerror}{\let\@latex@error\@gobble}
\makeatother
\usepackage{amsmath,amssymb,amsfonts}
\usepackage{graphicx}
\usepackage{caption}
\usepackage{subcaption}
\usepackage{textcomp}
\usepackage{xcolor}
\usepackage{longtable}
\usepackage{algorithmic}
\usepackage{bm}
\usepackage{diagbox}
\usepackage{booktabs}
\usepackage{url}
\usepackage{multirow}
\usepackage{algorithm}
\usepackage{algorithmic}
\usepackage{array}
\usepackage{textcomp}
\usepackage{stfloats}
\usepackage{url}
\usepackage{diagbox}
\usepackage{verbatim}
\usepackage{graphicx}
\usepackage{multirow}
\usepackage{xcolor}
\usepackage{amsthm}

\usepackage{stfloats}

\usepackage{tikz}
\usetikzlibrary{positioning,arrows.meta} 
\usepackage{graphicx}

\usepackage{bbm}

\def\BibTeX{{\rm B\kern-.05em{\sc i\kern-.025em b}\kern-.08em
    T\kern-.1667em\lower.7ex\hbox{E}\kern-.125emX}}

\usepackage{microtype}
\usepackage{float}

\title{ML-OPF-Bench: Benchmarking Machine Learning for Optimal Power Flow}

\author[1]{Xinyi Liu}
\author[2]{Xuan He}
\author[2]{Danny H.K. Tsang}
\author[3] {and Yize Chen}

\affiliation[1]{Dartmouth College}
\affiliation[2]{Hong Kong University of Science and Technology (Guangzhou)}
\affiliation[3]{University of Alberta}

\definecolor{verylightgray}{rgb}{0.9,0.9,0.9}
\definecolor{lightblue}{rgb}{0.733,0.875,1.0}
\definecolor{metablue}{rgb}{0, 0.392, 0.898}
\definecolor{forestgreen}{rgb}{0.208,0.667,0.235}
\definecolor{verylightblue}{rgb}{0.839,0.925,1.0}
\definecolor{veryverylightblue}{rgb}{0.92,0.96,1.0}

\abstract{
Machine Learning (ML) methods promise a fast solution process for Optimal Power Flow (OPF). While inconsistent test cases, implementations, and evaluation metrics across existing studies make it challenging to determine which algorithmic advances are most critical for real-world deployment. To this end, we propose ML-OPF-Bench, a unified benchmark for AC- and DC-OPF that evaluates representative ML algorithms under a consistent pipeline, stress-tests them across system sizes, distribution shifts, and resource budgets, and ranks them with a multi-objective framework. We find that prediction accuracy alone is not a reliable indicator of operational feasibility. Under heavily loaded, congested conditions, even the strongest in-distribution performers lose their advantage, while feasibility is maintained largely by post-processing that enforces the target constraints rather than by the underlying pure ML predictor. Data scaling shows that prediction accuracy and constraint violations follow different trajectories, whereas compute scaling shows that returns diminish and that larger models do not consistently perform better. These results expose critical trade-offs among ML mdethods' speed, accuracy, and feasibility, and offer practical guidance for future ML-OPF design. We open-source the benchmark as an extensible Python package for integrating new learning-based OPF algorithms and evaluating them under the same standard as the existing baselines.%
\footnote{%
\begin{tabular}[t]{@{}ll@{}}
Code: & \url{https://github.com/XinyiLiu04/ML-OPF-Bench} \\
Dataset: & \url{https://huggingface.co/datasets/xinyi-liu/ML-OPF-Bench}
\end{tabular}%
}}

\date{\today}
\correspondence{Yize Chen \email{yize.chen@ualberta.ca}}

\begin{document}

\maketitle

\section{Introduction}

The Optimal Power Flow (OPF) problem is one of the fundamental 
optimization tasks in power system operation and planning~\cite{khaloie2025review}, 
which is repeatedly solved to determine the most economical generation dispatch while satisfying physical and operational constraints throughout the network~\cite{Cain2012HistoryOO}. Modern power systems exhibit unprecedented complexity, making the 
efficient and reliable solution of OPF problems more challenging than ever.

The AC-OPF problem is NP-hard due to its nonconvex and nonlinear power flow equations~\cite{bienstock2019strong}. Traditional methods span mathematical programming such as sequential quadratic programming and interior-point~\cite{SQPSLP, interiorpoint}; metaheuristic algorithms~\cite{GA,swarm}; and convex relaxations such as semidefinite programming~\cite{SDP}. As these methods must be re-solved from scratch for every new load condition, they become slow and computationally burdensome on large networks or under high-frequency dispatch~\cite{frequency}, which often  forces operators to fall back on the linearized DC-OPF approximation, sacrificing 
accuracy and risking infeasibilities or suboptimalities~\cite{DCOPF}. 

As an alternative that reframes the trade-off between slow exact solvers and inexact approximations, many researchers have turned to Machine Learning (ML) to solve OPF problems. Among ML methods, Neural Networks (NNs) cast OPF as a learned mapping from operating conditions, such as the nodal active and reactive demand, to dispatch decisions, such as the generator active power and voltage magnitude setpoints, from which the remaining network state is recovered. By replacing per-instance optimization with a single forward pass, a trained network returns solutions orders of magnitude faster than iterative solvers~\cite{van2021machine}. Recent years have witnessed a rapidly expanding literature, with 
novel ML model architectures, training objectives, and constraint-handling strategies proposed 
in quick succession. With this proliferation, the pressing question is which of these design choices contributes most to OPF solution process in a reliable manner.

However, studies often adopt inconsistent test cases, sampling strategies, network sizes, and reported metrics~\cite{Deka2019LearningFD,Qpost,DeepDCOPF,DeepACOPF,GNN,lagrangian,KKTAC,KKTDC,chen2022learning,rlopf,linear}. In addition, many proposed algorithms lack open-source implementations or sufficient experimental detail to reproduce their results, and this field still lacks a shared infrastructure for systematic benchmarking. Though recent efforts have tried to bridge the data gap for learning-based OPF solvers~\cite{joswig2022opf,shen2026proopf}, reported gains still cannot be compared or verified on a common scale. Moreover, it remains unclear whether performance carries over to out-of-distribution operating points, particularly the heavily loaded, congested regimes common in practice. Such conditions are rarely included in existing evaluations, leaving operational feasibility and robustness underinvestigated.

\begin{figure*}
    \centering
    \includegraphics[width=0.8\linewidth]{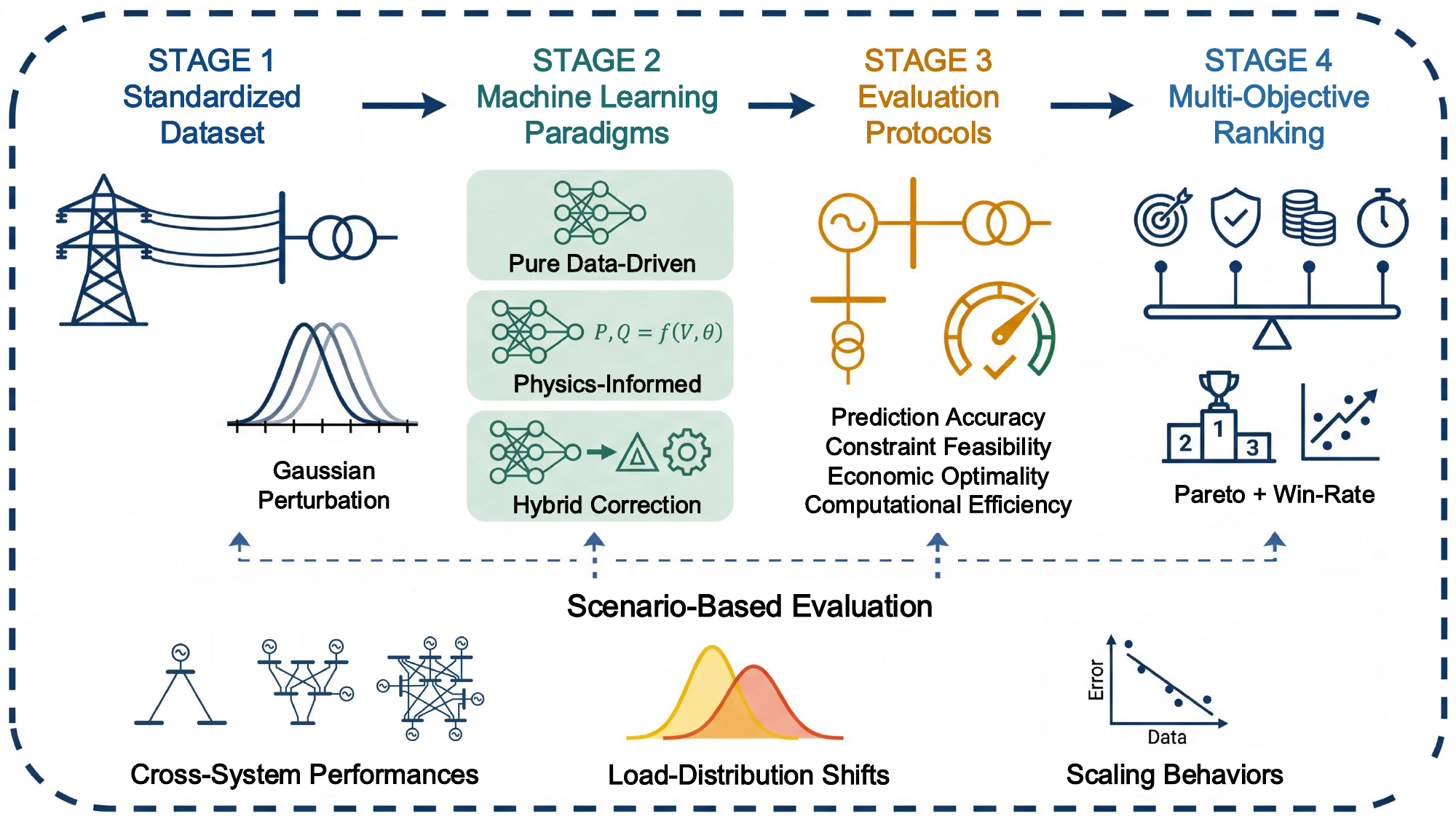}
    \caption{Structural overview of the ML-OPF Benchmark.}
    \label{fig:workflow}
\end{figure*}

To bridge these gaps, we present ML-OPF-Bench, a framework for unified,
reproducible, and comparable evaluation of ML methods in DC- and AC-OPF,
whose overall architecture is illustrated in Fig.~\ref{fig:workflow}. Solutions are validated with PyPower
    power flow~\cite{Zimmerman1997MATPOWER} for AC and PTDF-based
    computation for DC, and every method is evaluated on a common set of
    metrics covering prediction accuracy, constraint feasibility, economic
    optimality, and computational efficiency. Our major
contributions are as follows.

\begin{itemize}
    \item \textbf{Standardized and reproducible benchmark.}
We develop ML-OPF-Bench, a unified framework for comparing
data-driven, physics-informed, and hybrid methods for
AC- and DC-OPF.
Methods use common IEEE test systems~\cite{PGlib},
load distributions~\cite{Gaussian}, and data splits,
and are evaluated using shared metrics covering accuracy,
feasibility, optimality, and computational efficiency.
Pareto levels and pairwise win rates summarize their
multi-objective performance.
We release the framework as an extensible
Python package with common interfaces for adding
methods and test systems.

\item \textbf{Scenario-guided evaluation protocol.} Beyond in-distribution comparison, we evaluate performance across system sizes, wider load distributions, and
heavily loaded, congested conditions. We also examine how prediction accuracy and constraint
performance change with training-data volume, model
capacity, and computational budget.

\item \textbf{Empirical insights.} Prediction accuracy alone is not a reliable
indicator of operational readiness. Near-optimal cost can coexist with
constraint violations, as when linear regression attains a cost gap of only
$-0.003\%$ on 300-bus DC system while still violating network limits, yet
this simple baseline remains competitive overall. Under heavily loaded,
congested conditions, even the strongest in-distribution performers degrade
substantially, with the prediction error of the top-ranked AC method,
\texttt{DNN}, rising from 0.306\% to 87.063\% on 118-bus. In this
regime, feasibility is maintained largely by post-processing that enforces the
target constraints rather than by the learning formulation itself, at the cost of
additional online computation. These limitations persist as training resources
grow. More data and compute generally improve accuracy with diminishing
returns, while gains from larger models are uneven, and a four-layer AC
network of width 512 is less accurate than one of width 128 despite nearly
five times the compute. In the data-scaling study, accuracy gains do not consistently translate into lower constraint violations, particularly for AC reactive power~\cite{Liu2026ScalingLO}.
\end{itemize}

\begin{table*}[htbp]
\centering
\caption{Overview of benchmarked methods and their core characteristics.}
\label{tab:method_overview}
\footnotesize
\setlength{\tabcolsep}{4pt}
\renewcommand{\arraystretch}{0.85}
\begin{tabular}{llll}
\toprule
\textbf{Paradigm}
& \textbf{Method}
& \textbf{Training Objective}
& \textbf{Constraint Handling} \\
\midrule
\multirow{3}{*}{Pure Data-Driven}
&  LR~[\citenum{linear}]
& Least squares
& No explicit constraint penalties \\
& DNN~[\citenum{lagrangian,DeepDCOPF,DeepACOPF}]
& MSE
& No explicit constraint penalties \\
& GNN~[\citenum{GNN}]
& MSE
& Graph features; no constraint penalties \\
\midrule
\multirow{6}{*}{Physics-Informed}
& Constraint Penalty (CP)~[\citenum{DeepDCOPF,DeepACOPF}]
& MSE + violation penalties
& Fixed-weight penalties in loss \\
& Multiplier Updates (MU)~[\citenum{lagrangian}]
& MSE + weighted violations
& Adaptive multipliers on violations \\
& KKT Embedded (KKT)~[\citenum{KKTAC,KKTDC}]
& Primal/dual + KKT residuals
& Optimality residuals in loss \\
& NGT~[\citenum{unsupervised}]
& Cost + constraint penalties
& Physics-based penalties \\
& E-NGT~[\citenum{unsupervised}]
& Cost + penalties + supervised loss
& Physics-based penalties \\
& RL~[\citenum{rlopf}]
& Cost-and-violation reward
& Reward penalties \\
\midrule
\multirow{4}{*}{Hybrid}
& QP Project-DC (QP)~[\citenum{DeepDCOPF}]
& MSE + violation penalties
& Projection onto DC feasible region \\
& Q Correction-AC (QC)~[\citenum{Qpost}]
& MSE
& PV-to-PQ switching and PF re-solve \\
& Feasibility Recovery-AC (FR)~[\citenum{DeepACOPF}]
& MSE + violation penalties
& Violation screening and OPF recovery \\
& Active Set (AS)~[\citenum{Deka2019LearningFD}]
& Cross-entropy
& Active-set prediction and LP recovery \\
\bottomrule
\end{tabular}
\end{table*}

By providing a transparent and extensible benchmarking platform, ML-OPF-Bench enables the community to effectively develop, compare, and improve ML-based OPF solutions, accelerating progress toward deployable, reliable tools that complement traditional optimization methods in power system operations. The remainder of this paper is organized as follows. Section~\ref{sec: framework} introduces the standard OPF problems, benchmarking pipelines, and benchmarked algorithms. Section~\ref{sec: setup} presents dataset generation procedure and experimental settings. Section~\ref{sec:results} reports the results across all evaluation scenarios and discusses the key findings and trade-offs. Finally, Section~\ref{sec:conclusion} concludes the paper and outlines directions for future research.

\section{Benchmark Framework}
\label{sec: framework}
\subsection{OPF Formulations}

\begin{table}[h]
\centering
\caption{Notations for ML-OPF-Bench.}
\label{tab:notation}
\small
\begin{tabular}{ll}
\toprule
\textbf{Notation} & \textbf{Definition} \\
\midrule
$\mathcal{G}, \mathcal{G}_{\text{ns}}$ & Set of generators/non-slack generators \\
$\mathcal{B}$ & Set of buses \\
$\mathcal{L}$ & Set of transmission lines \\
$N_g, N_b, N_l, N_d$ & Number of generators/buses/lines/loads \\
$N$ & Number of evaluation samples \\
$P_g, Q_g$ & Active/reactive power generation \\
$P_d, Q_d$ & Active/reactive power demand \\
$V, \theta$ & Voltage magnitude/angle \\
$G_{ij}, B_{ij}$ & Conductance/susceptance of branch $(i,j)$ \\
$P_{ij}^f, Q_{ij}^f$ & Active/reactive power flows on branch $(i,j)$ \\
$S_{ij}$ & Apparent (AC) or active (DC) power flow \\
$\theta_{ij}^{\Delta}$ & Maximum allowed angle difference \\
$\theta_{\mathrm{ref}}$ & Reference (slack) bus angle \\
$c_{2,i}, c_{1,i}, c_{0,i}$ & Generation cost coefficients \\
$C(P_g)$ & Cost function \\
$\hat{\cdot}$ & Predicted quantity \\
$(\cdot)^{*}$ & Ground-truth / solver-optimal quantity \\
$N_h$ & Number of hidden layers \\
$N_n$ & Number of neurons per hidden layer \\
\bottomrule
\end{tabular}
\end{table}

AC-OPF seeks to determine the optimal generator dispatch that minimizes the total generation cost while satisfying physical and operational constraints~\cite{Frank2016IntroductionOPF}, detailed in Model 1.

\begin{figure}[t]
\hrule height 0.8pt
\vspace{4pt}
\noindent\textbf{Model 1: AC Optimal Power Flow (AC-OPF)}
\vspace{2pt}
\hrule height 0.4pt
\vspace{4pt}
\begingroup
\small
\begin{subequations}\label{eq:acopf}
\begin{flalign}
\mathcal{O}(P_d, Q_d) = \mathop{\min}_{P_g, Q_g, V, \theta} \sum_{i=1}^{N_g} \left( c_{2,i} P_{g,i}^2 + c_{1,i} P_{g,i} + c_{0,i} \right) &&
\label{eq:acobj}
\end{flalign}
\vspace{-14pt}
\noindent\textbf{subject to:}
\vspace{8pt}
\addtolength{\jot}{0.3em}
\begin{align}
& P_{g,i} - P_{d,i} = V_i \sum_{j\in\mathcal{B}} V_j ( G_{ij}\cos\theta_{ij} + B_{ij}\sin\theta_{ij} ), \forall i\in\mathcal{B} \label{eq:pbal}\\
& Q_{g,i} - Q_{d,i} = V_i \sum_{j\in\mathcal{B}} V_j ( G_{ij}\sin\theta_{ij} - B_{ij}\cos\theta_{ij} ), \forall i\in\mathcal{B} \label{eq:qbal}\\
& P_{g,i}^{\min} \leqslant P_{g,i} \leqslant P_{g,i}^{\max}, \hfill \forall i\in\mathcal{G} \label{eq:pglim}\\
& Q_{g,i}^{\min} \leqslant Q_{g,i} \leqslant Q_{g,i}^{\max}, \hfill \forall i\in\mathcal{G} \label{eq:qglim}\\
& V_i^{\min} \leqslant V_i \leqslant V_{i}^{\max}, \hfill \forall i\in\mathcal{B} \label{eq:vlim}\\
& -\theta_{ij}^{\Delta} \leqslant \theta_i - \theta_j \leqslant \theta_{ij}^{\Delta}, \hfill \forall (i,j)\in\mathcal{L} \label{eq:angdiff}\\
& (P_{ij}^f)^2 + (Q_{ij}^f)^2 \leqslant (S_{ij}^{\max})^2, \hfill \forall (i,j)\in\mathcal{L} \label{eq:slim}\\
& \theta_{\mathrm{ref}} = 0 \label{eq:refbus}
\end{align}
\end{subequations}
\endgroup
\hrule height 0.8pt
\end{figure}

DC-OPF is a linearized approximation of AC-OPF~\cite{DCOPF}, where voltage magnitudes are fixed at 1.0 p.u., angle differences across branches are small, and line resistances are negligible compared to reactances. In DC, the active power balance simplifies to:
\begin{equation}\label{eq:dcpbal}
P_{g,i} - P_{d,i} = \sum_{j \in \mathcal{B}} B_{ij} (\theta_i - \theta_j), \quad \forall i \in \mathcal{B}
\end{equation}
The nonlinear apparent power flow constraint~\eqref{eq:slim} is replaced by a linear active power flow limit:
\begin{equation}\label{eq:dcflowlim}
-P_{ij}^{\max} \leqslant B_{ij}(\theta_i - \theta_j) \leqslant P_{ij}^{\max}, \quad \forall (i,j) \in \mathcal{L}
\end{equation}

Table~\ref{tab:notation} summarizes notations used in this paper.

\subsection{Benchmarking Pipeline}

The benchmark distinguishes model prediction and additional feasibility correction. Additional projection and correction procedures are identified
explicitly in the hybrid methods.
Table~\ref{tab:method_overview} summarizes the benchmarked
methods and their abbreviations.

\noindent\textbf{DC-OPF pipeline.}
Methods map active loads to generator dispatch through direct
prediction or active-set recovery.
Most predictors use MinMax-scaled loads, while \texttt{LR}
uses unscaled per-unit loads and \texttt{RL} uses standardized
loads.
For \texttt{GNN}, bus-level inputs additionally include a
generator indicator, aggregated generation bounds, normalized
node degree, and mean neighboring load.
When only non-slack generation is predicted, slack generation
is reconstructed from system-wide power balance.
Line flows are computed from the resulting nodal injections
using the PTDF matrix.

\noindent\textbf{AC-OPF pipeline.}
The basic pipeline predicts non-slack active generation and
generator-bus voltage magnitudes,
$(\hat{P}_g^{\mathrm{ns}},\hat{V}_{\mathrm{gen}})$,
and uses these setpoints in a Newton--Raphson power flow
to obtain the remaining variables.
\texttt{LR}, \texttt{DNN}, and \texttt{CP} predict both
quantities; the predictors used by \texttt{QC} and
\texttt{FR} follow the same setpoint representation.

Several methods produce a broader state.
\texttt{MU} predicts non-slack active generation, reactive
generation at all generators, and voltage magnitudes and
angles at all buses.
\texttt{KKT} predicts active and reactive generation at all
generators, rectangular voltages at all buses, and dual
variables; voltage magnitudes are obtained from the
rectangular components.
\texttt{NGT} and \texttt{E-NGT} predict voltage magnitudes
and angles at non-zero-injection buses, recover the remaining
voltages through Kron reduction~\cite{kron}, and calculate
generation from the nodal injection equations.
For these four methods, only the resulting
$(\hat{P}_g^{\mathrm{ns}},\hat{V}_{\mathrm{gen}})$
enter the common power-flow evaluator.
Reactive generation, bus angles, PQ-bus voltage magnitudes,
and slack active generation are then recomputed by power flow.
Direct outputs of \texttt{MU} and \texttt{KKT} are examined
separately in Appendix~\ref{app:direct}.

For methods supplying only active-generation predictions
or recovered dispatches, generator voltage-magnitude
setpoints come from the case data~\cite{PGlib} for
\texttt{RL}, a preliminary power-flow state for
\texttt{GNN}, and training-set means for \texttt{AS}.
The preliminary \texttt{GNN} state is obtained by solving
DC-OPF followed by AC power flow, and also supplies its
bus-level input features.

\subsection{Benchmarked Algorithms}
\label{sec:algorithms}

We compare three families according to how they learn the
dispatch map and incorporate constraints.
Detailed equations and implementation adaptations are
provided in Appendix~\ref{app:methods}.

\noindent\textbf{Pure Data-Driven Scheme.}
These methods use supervised regression without explicit
constraint penalties in the training loss.
\texttt{LR} fits a multi-output linear least-squares
map~\cite{linear}.
\texttt{DNN} uses a feedforward network trained with
mean squared error (MSE).
\texttt{GNN} applies Chebyshev graph convolutions to
bus-level features~\cite{GNN}.

\noindent\textbf{Physics-Informed Scheme.}
\texttt{CP} augments supervised regression with
constraint-related penalty terms~\cite{DeepDCOPF,DeepACOPF}.
\texttt{KKT} combines state and multiplier supervision
with optimality residuals~\cite{KKTAC,KKTDC}, while
\texttt{MU} updates multipliers associated with
physical penalty terms~\cite{lagrangian}.
\texttt{NGT} learns from generation cost and physical
residuals, and \texttt{E-NGT} adds supervised updates
using labeled samples~\cite{unsupervised}.
\texttt{RL} learns a one-step dispatch policy using DDPG
with a cost-and-violation reward~\cite{rlopf}.

\noindent\textbf{Hybrid Scheme.}
These methods combine learned predictions with explicit post
optimization or constraint-handling procedures.
DC \texttt{QP} projects the \texttt{CP} prediction onto
the implemented linear feasible region~\cite{DeepDCOPF}.
AC \texttt{QC} applies reactive-limit correction through
PV-to-PQ switching and a further power-flow solve~\cite{Qpost}.
AC \texttt{FR} attempts OPF recovery when the
\texttt{CP} power-flow result fails a constraint
screen~\cite{DeepACOPF}.
\texttt{AS} predicts active-set classes and uses LP-based
dispatch recovery and candidate selection~\cite{Deka2019LearningFD}.

\section{Benchmark Setup}
\label{sec: setup}

\textbf{Datasets.}
We benchmark on the IEEE 30-, 118-, and 300-bus cases from
PGLib-OPF v23.07~\cite{PGlib}.
Using the Julia package PowerModels.jl~\cite{powermodeljl}, AC datasets are generated with IPOPT~\cite{Wachter2006interior} and DC datasets with its default linear programming interface.
For each case, 50,000 load are attempted~\cite{DeepDCOPF,joswig2022opf}, retaining only converged and
feasible solutions.
Loads are perturbed using Gaussian noise~\cite{Gaussian} with different variance depending on scenarios. All data are stored in per-unit with base MVA. Table~\ref{tab:ieee_test_cases} summarizes the system sizes, hidden-layer widths, and batch sizes.

\begin{table}[htbp]
   \centering
   \caption{Test-system sizes and network configurations.}
   \label{tab:ieee_test_cases}
   \footnotesize
   \setlength{\tabcolsep}{4pt}
   \renewcommand{\arraystretch}{0.85}
   \begin{tabular}{lccccccc}
     \toprule
     Test Case & $N_b$ & $N_g$ & $N_d$ & $N_l$
     & $N_n$(AC) & $N_n$(DC) & Batch \\
     \midrule
     30-bus  & 30  & 6  & 21  & 41
     & 64/32   & 32/16   & 32 \\
     118-bus & 118 & 54 & 99  & 186
     & 256/128 & 128/64  & 64 \\
     300-bus & 300 & 69 & 191 & 411
     & 512/256 & 256/128 & 128 \\
     \bottomrule
   \end{tabular}
\end{table}

\textbf{Settings.}
Neural models are implemented in PyTorch~\cite{Paszke2019PyTorchAI}.
The baseline feedforward configurations use two hidden layers
with the widths listed in Table~\ref{tab:ieee_test_cases}.
These widths provide common reference configurations, while
method-specific structures are retained, including the
multi-branch architectures of \texttt{MU} and \texttt{KKT},
the graph-convolution layers of \texttt{GNN}, and the
actor--critic structure of \texttt{RL}.
The results characterize the specified implementations and unified
training protocol rather than the best performance achievable
through exhaustive per-method tuning.

Validation-based early stopping is used where enabled by the
training protocol. \texttt{MU} is trained for 80 epochs~\cite{lagrangian}, while
\texttt{NGT}/\texttt{E-NGT} for 2,500 epochs
on the 30- and 300-bus systems and 3,000 epochs on the 118-bus
system~\cite{unsupervised}. \texttt{RL} uses DDPG, with checkpoint selection based
on validation reward. Neural training uses
Adam~\cite{Adam} with a learning rate of 0.001,
except for \texttt{RL}, which uses 0.0003.
Neural training runs on NVIDIA A10 GPUs, while power-flow and linear regression run on Intel Xeon Platinum 8358 CPUs. As the CPU-bound power flow remains a bottleneck for computational speedup, developing GPU-native differentiable power flow solvers represents an interesting direction to fully exploit hardware acceleration in ML-OPF~\cite{bienstock2019strong}.

\textbf{Metrics.}
Following metrics are adopted for evaluation.
For AC-OPF, metrics are only computed over samples with
converged power-flow results. 

\begin{itemize}

\item \textbf{Prediction Accuracy.}
Mean absolute error (MAE) is normalized by the mean absolute
ground-truth:
\begin{equation}
\small 
\operatorname{MAE}(X)
=
\frac{\operatorname{mean}(|\hat{X}-X^*|)}
     {\operatorname{mean}(|X^*|)}
\times 100\%,
X\in\{P_g^{\mathrm{ns}},V_{\mathrm{gen}}\}.
\end{equation}
Active-generation MAE covers non-slack generators;
voltage-magnitude MAE is reported for AC and covers generator
buses.

\item \textbf{Constraint Violations.}
We use a mean-of-max aggregation: the largest component
violation is computed for each sample and then averaged over
samples. For non-slack active generation (in p.u.),
\begin{equation}
\small 
\label{Eq: ViolPg}
\operatorname{Viol}_{\mathrm{Pg}}
=
\frac{1}{N}\sum_{i=1}^{N}
\max_{j\in\mathcal{G}_{\mathrm{ns}}}
\left[
\max\left(
\hat{P}_{g,j}^{(i)}-P_{g,j}^{\max},
P_{g,j}^{\min}-\hat{P}_{g,j}^{(i)},
0
\right)
\right].
\end{equation}
The slack violation $\text{Viol}_{\text{Pg}}^0$ is reported separately to assess power balance. For AC-OPF, analogous definitions give $\text{Viol}_{\text{Qg}}$ over all generators and $\text{Viol}_{\text{V}}$ over all buses.

Thermal violations are normalized by branch
ratings:
\begin{equation}
\label{Eq: Viollf}
\operatorname{Viol}_{\mathrm{lf}}
=
\frac{1}{N}\sum_{i=1}^{N}
\max_{e\in\mathcal{E}}
\left[
\frac{|\hat{S}_{e}^{(i)}|}{S_e^{\max}}-1
\right]_+,
\end{equation}
where $[x]_+=\max(x,0)$.
For AC, $\mathcal{E}$ includes both ends of each rated branch
and $|\hat{S}_e|$ is apparent-power magnitude.
For DC, the metric uses absolute active-power flow. A value of 0.01 corresponds to a 1\% overload.

\item \textbf{Economic Optimality.}
The cost gap compares the evaluated dispatch with the
reference OPF solution:
\begin{equation}
\operatorname{Gap}_{\mathrm{cost}}
=
\frac{1}{N}\sum_{i=1}^{N}
\frac{
C(\hat{P}_g^{(i)})-C(P_g^{*(i)})
}{
C(P_g^{*(i)})
}
\times 100\%.
\end{equation}
A negative cost gap does not by itself indicate a better
feasible solution, since an infeasible dispatch may have
a lower generation cost.

\item \textbf{Computational Efficiency.}
We report training wall-clock time in seconds and mean
inference time in milliseconds per sample.
The inference times cover the complete batch-one prediction
pipeline, including preprocessing, prediction, reconstruction,
and method-specific post optimization or correction.
For AC \texttt{GNN}, this includes the DC-OPF and preliminary
AC power flow used to generate its input features.

\end{itemize}

\textbf{Rankings.}
Ranking is a familiar device in machine learning benchmarking, where methods are commonly compared and ordered along a shared performance metric~\cite{meta,correctbench}. We adopt this device but adapt it to the particular demands of OPF. Unlike many learning tasks, OPF is a physics-constrained problem, and a dispatch decision is only actionable if it also satisfies generator, voltage, and thermal limits within a practical time budget. To this end, we rank algorithms along four objectives that mirror the evaluation metrics defined above:
$\operatorname{MAE}_{\mathrm{Pg}}$,
$|\operatorname{Cost}_{\mathrm{gap}}|$,
$\operatorname{Viol}_{\mathrm{total}}$, and inference time. The aggregate violation metric is an equal-weight sum of
the reported components. For DC, the violation is
\begin{equation}
\operatorname{Viol}_{\mathrm{total}}^{\mathrm{DC}}
=
\operatorname{Viol}_{\mathrm{Pg}}
+
\operatorname{Viol}_{\mathrm{Pg}}^{0}
+
\operatorname{Viol}_{\mathrm{lf}},
\end{equation}
while for AC-OPF it additionally includes $\text{Viol}_{\text{Qg}}$ and $\text{Viol}_{\text{V}}$ terms. Within each system, a method $a$ dominates method $b$ if
it is no worse on every objective and strictly better on
at least one. Non-dominated sorting assigns Pareto
levels, with level 1 identifying the non-dominated front. We additionally compute pairwise win rates~\cite{fevbench}.
Each system--objective pair forms one comparison unit,
giving $U=3\times4=12$ units.
Let $m_{uj}$ denote the value of method $j$ on
unit $u$. Its pairwise win rate against method $k$ is:
\begin{equation}
W_{jk}
=
\frac{1}{U}\sum_{u=1}^{U}
\left[
\mathbf{1}(m_{uj}<m_{uk})
+
\frac{1}{2}\, \mathbf{1}(m_{uj}=m_{uk})
\right],
\label{eq:winrate}
\end{equation}
where ties receive half credit.
The marginal win rate is:
\begin{equation}
W_j
=
\frac{1}{M-1}\sum_{k\ne j}W_{jk},
\label{eq:marginalwin}
\end{equation}
where $M$ is the number of methods compared within the formulation. Methods are ordered by ascending mean Pareto level across the three systems, with ties broken by descending marginal win rate.

\textbf{Scenarios.}
Our evaluation scenarios are organized around the deployment challenges a learned OPF
solver must withstand. In practice, a solver intended for real grid operational tasks must hold its accuracy and feasibility as the network grows from a small test system to a larger transmission grid~\cite{crosssystems}. Real systems rarely sit at their nominal load, so in-distribution standing must carry over once the operating point moves toward larger load variance or heavier and more congested demand~\cite{renewable}. Labeled OPF solutions are also expensive to generate~\cite{unsupervised}, which makes the amount of data each paradigm requires a practical concern. Finally, training compute is limited as well, so it matters whether larger models and longer training buy proportionally better accuracy~\cite{Kaplan2020ScalingLF,Hoffmann2022TrainingCL}.

\begin{itemize}

\item \textbf{Cross-System Performance.}
Methods are trained and evaluated independently on the 30-, 118-, and 300-bus systems.
For each formulation and system, 12,000 samples are drawn and divided into 10,000 training, 1,000 validation, and 1,000 test samples. Methods share the same split within each setting.

\item \textbf{Generalization under Distribution Shifts.}
Models trained on the 118-bus typical-condition (TYP)
dataset are evaluated without retraining under two
test-time shifts.
The larger-variance scenario increases the relative standard deviation of the load perturbations from 0.12 to 0.25 while retaining the original operating
limits. The Active Power Increase (API) scenario uses the loading conditions and operating limits of the corresponding PGLib API case~\cite{PGlib}, representing a heavily loaded, congested setting with a changed feasible region. Figure~\ref{fig:Gaussian} illustrates the load distributions.

\begin{figure}[htbp]
    \centering
    \includegraphics[
        trim={5cm 3.5cm 5cm 3.5cm},
        clip,
        width=\linewidth
    ]{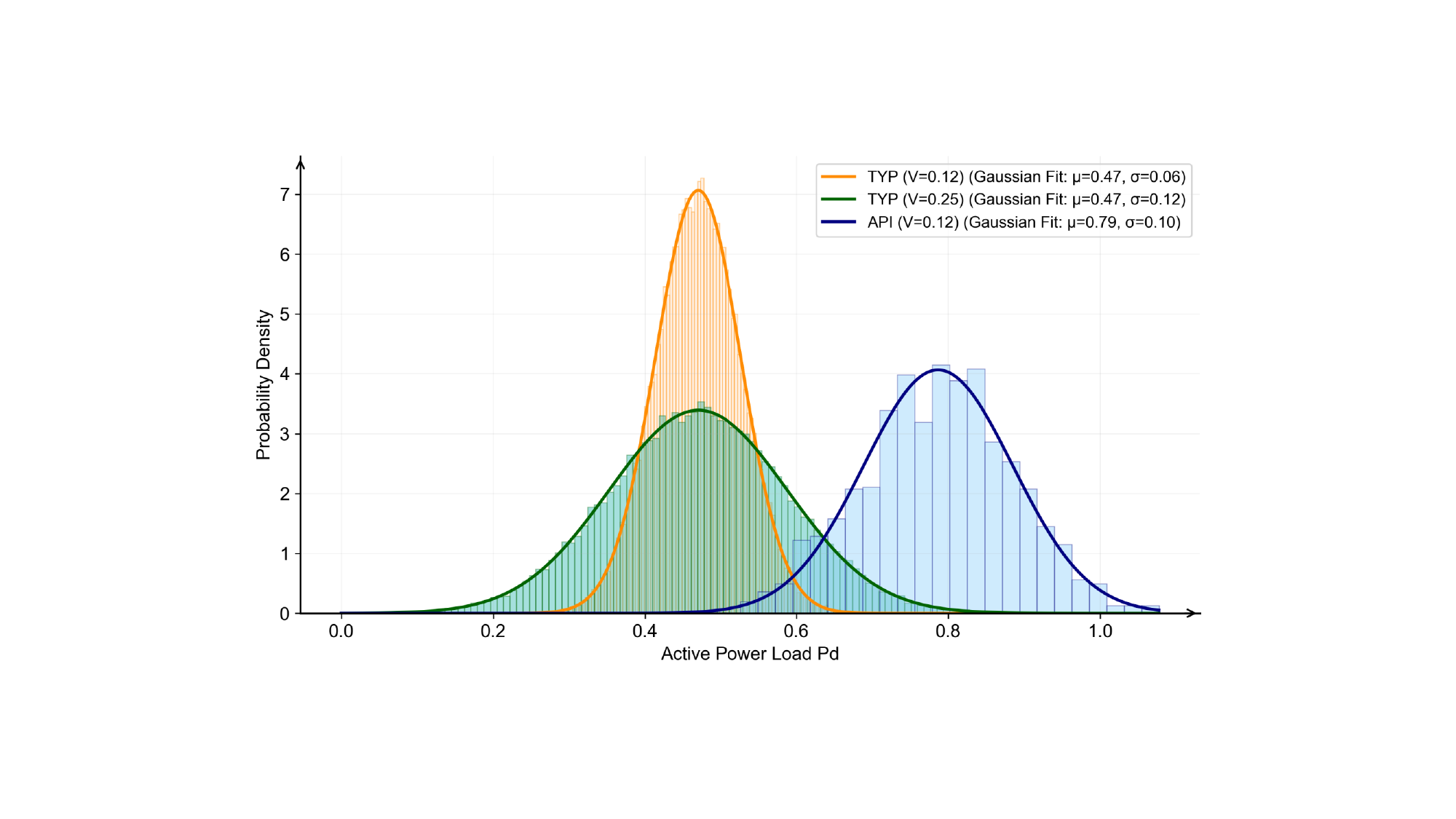}
    \caption{Active-load distributions at Bus 12 of the
    IEEE 118-bus system under typical conditions,
    larger load variance, and the API scenario.}
    \label{fig:Gaussian}
\end{figure}

\item \textbf{Training-Data Scaling.}
On the 118-bus system, we reserve 10\% of the available
dataset for validation and another 10\% for testing,
and draw training subsets of $N\in\{1,5,12,20,30,35\}\mathrm{K}$
from the remaining samples. All training sizes therefore share the same held-out data. Inspired by AI compute-accuracy-data scaling laws, to characterize dataset size's impacts on ML-OPF performance, in this work we fit
power-law function of the following form~\cite{hestness2017deep,Kaplan2020ScalingLF}:
\begin{equation}
\label{eq:power law}
m(N)=A(N/N_0)^\alpha,
\qquad N_0=10{,}000.
\end{equation}
Parameters are estimated by ordinary least squares
in log space.
For lower-is-better metrics, a negative exponent
indicates a decreasing trend.
$R^2_{\log}$ measures the goodness of fit on the log-transformed metric values. 

From the same runs, we also examine the recorded validation-loss
trajectories of the iterative methods to characterize how training
progress changes with sample size, see Appendix~\ref{app:data-scaling} for more details.

\item \textbf{Model Capacity and Compute Budgets.}
We fix the 118-bus training set at 12,000 samples and
vary neural network depth over $\{2,3,4\}$ hidden layers
and width over $\{64,128,256,512\}$ neurons per layer.
These experiments retain early stopping and use the
same validation and test sets as the data-scaling study. Compute is measured as cumulative algorithmic floating-point
operations (FLOPs) for forward propagation, backpropagation,
Adam updates, and validation. A multiply--accumulate counts as two FLOPs~\cite{Hoffmann2022TrainingCL}. The detailed operation counts
are provided in Appendix~\ref{app:compute}.

\end{itemize}

\section{Benchmark Results and Analysis}
\label{sec:results}
\begin{table*}[h]
\centering
\caption{Performance of DC-OPF methods across test systems.}
\label{tab:main-dc}
\label{tab:dc_stats}
\scriptsize
\setlength{\tabcolsep}{3pt}
\resizebox{\textwidth}{!}{%
\begin{tabular}{ll*{12}{r}}
\toprule
& & \multicolumn{4}{c}{30-bus}
& \multicolumn{4}{c}{118-bus}
& \multicolumn{4}{c}{300-bus} \\
\cmidrule(lr){3-6}
\cmidrule(lr){7-10}
\cmidrule(lr){11-14}
Category & Method
& MAE$_{\mathrm{Pg}}$ & Viol$_{\mathrm{total}}$ & Cost gap & Infer
& MAE$_{\mathrm{Pg}}$ & Viol$_{\mathrm{total}}$ & Cost gap & Infer
& MAE$_{\mathrm{Pg}}$ & Viol$_{\mathrm{total}}$ & Cost gap & Infer \\
\midrule
\multirow{3}{*}{\shortstack[l]{Data-Driven}}
& LR    & 0.000 & 0.000 & 0.000  & 0.119 & 1.013 & 0.076 & -0.004 & 0.121 & 2.413 & 0.490 & -0.003 & 0.128 \\
& DNN   & 0.162 & 0.000 & 0.040  & 0.454 & 0.711 & 0.030 & 0.001  & 0.439 & 1.789 & 0.991 & -0.152 & 0.456 \\
& GNN   & 1.369 & 0.003 & -0.034 & 2.950 & 1.073 & 0.021 & -0.003 & 2.906 & 3.171 & 1.347 & 0.015  & 3.008 \\
\midrule
\multirow{6}{*}{\shortstack[l]{Physics-Informed}}
& CP    & 1.118  & 0.003 & -0.109  & 0.448 & 0.364 & 0.005 & 0.011  & 0.451 & 1.170  & 0.550 & 0.019  & 0.481 \\
& MU    & 1.134  & 0.003 & -0.079  & 0.463 & 0.377 & 0.004 & 0.014  & 0.483 & 1.925  & 0.185 & 0.928  & 0.473 \\
& KKT   & 0.058  & 0.000 & -0.015  & 0.438 & 0.838 & 0.048 & -0.040 & 0.440 & 1.540  & 0.967 & -0.071 & 0.447 \\
& NGT   & 0.420  & 0.002 & -0.128  & 0.383 & 0.994 & 0.000 & 0.075  & 0.388 & 4.586  & 0.018 & 0.264  & 0.384 \\
& E-NGT & 0.225  & 0.001 & -0.069  & 0.379 & 3.057 & 0.003 & 0.574  & 0.415 & 1.697  & 0.021 & 0.878  & 0.388 \\
& RL    & 80.187 & 0.327 & -24.500 & 0.560 & 1.228 & 0.024 & -0.028 & 0.558 & 15.920 & 0.736 & -4.811 & 0.545 \\
\midrule
\multirow{2}{*}{Hybrid}
& QP    & 0.373 & 0.000 & 0.113 & 0.973 & 0.421 & 0.000 & 0.092  & 4.511 & 1.333  & 0.000 & 0.564 & 11.328 \\
& AS    & 0.000 & 0.000 & 0.000 & 4.442 & 0.653 & 0.019 & -0.399 & 4.608 & 26.294 & 3.283 & 3.285 & 5.790 \\
\bottomrule
\end{tabular}%
}
\end{table*}
 
\begin{table*}[h]
\centering
\caption{Performance of AC-OPF methods across test systems.}
\label{tab:main-ac}
\label{tab:ac_stats}
\scriptsize
\setlength{\tabcolsep}{3pt}
\resizebox{\textwidth}{!}{%
\begin{tabular}{ll*{15}{r}}
\toprule
& & \multicolumn{5}{c}{30-bus}
& \multicolumn{5}{c}{118-bus}
& \multicolumn{5}{c}{300-bus} \\
\cmidrule(lr){3-7}
\cmidrule(lr){8-12}
\cmidrule(lr){13-17}
Category & Method
& MAE$_{\mathrm{Pg}}$ & MAE$_V$
& Viol$_{\mathrm{total}}$ & Cost gap & Infer
& MAE$_{\mathrm{Pg}}$ & MAE$_V$
& Viol$_{\mathrm{total}}$ & Cost gap & Infer
& MAE$_{\mathrm{Pg}}$ & MAE$_V$
& Viol$_{\mathrm{total}}$ & Cost gap & Infer \\
\midrule
\multirow{3}{*}{\shortstack[l]{Data-Driven}}
& LR
& 0.040 & 0.005 & 0.001 & 0.000 & 9.720
& 0.518 & 0.066 & 0.093 & -0.002 & 17.921
& 4.565 & 0.248 & 0.572 & -0.028 & 26.831 \\
& DNN
& 0.240 & 0.015 & 0.006 & 0.018 & 9.621
& 0.306 & 0.041 & 0.076 & -0.012 & 11.060
& 3.995 & 0.216 & 0.451 & 0.701 & 17.967 \\
& GNN
& 3.245 & 3.576 & 0.553 & 0.514 & 22.332
& 0.958 & 3.702 & 1.295 & 0.403 & 33.259
& 6.566 & 3.924 & 4.791 & 0.062 & 68.736 \\
\midrule
\multirow{6}{*}{\shortstack[l]{Physics-Informed}}
& CP
& 1.970 & 0.253 & 0.014 & -0.086 & 9.666
& 1.088 & 0.122 & 0.078 & -0.426 & 11.267
& 10.718 & 0.797 & 0.686 & 1.011 & 17.868 \\
& MU
& 0.569 & 0.032 & 0.005 & -0.121 & 9.973
& 1.341 & 0.074 & 0.118 & 0.078 & 11.663
& 8.067 & 0.595 & 2.692 & 0.029 & 18.957 \\
& KKT
& 0.624 & 0.027 & 0.011 & -0.136 & 9.683
& 0.669 & 0.107 & 0.115 & -0.011 & 11.658
& 4.150 & 0.522 & 1.128 & 0.033 & 18.275 \\
& NGT
& 102.569 & 3.853 & 1.459 & -30.545 & 10.641
& 13.976 & 3.328 & 0.503 & 1.482 & 12.413
& 102.907 & 4.366 & 17.921 & 5.855 & 19.322 \\
& E-NGT
& 32.507 & 0.105 & 0.115 & -5.565 & 10.903
& 15.509 & 0.396 & 0.483 & 1.500 & 12.489
& 15.155 & 0.475 & 2.899 & 1.152 & 19.283 \\
& RL
& 61.060 & 3.576 & 0.955 & -18.478 & 9.736
& 14.718 & 3.702 & 1.325 & 3.817 & 11.183
& 27.523 & 3.919 & 4.113 & 4.725 & 17.799 \\
\midrule
\multirow{3}{*}{Hybrid}
& QC
& 0.378 & 0.343 & 0.450 & 0.227 & 18.200
& 0.272 & 0.065 & 0.073 & 0.000 & 21.702
& 3.205 & 0.291 & 0.305 & 0.291 & 36.565 \\
& FR
& 0.000 & 0.001 & 0.000 & 0.000 & 232.519
& 0.000 & 0.001 & 0.000 & 0.000 & 563.989
& 1.175 & 0.088 & 0.074 & 0.117 & 1553.091 \\
& AS
& 20.187 & 0.068 & 0.042 & 6.654 & 11.005
& 8.403 & 0.234 & 0.308 & 0.478 & 15.198
& 26.604 & 0.442 & 16.072 & -5.625 & 36.817 \\
\bottomrule
\end{tabular}%
}
\par\smallskip
\begin{minipage}{\textwidth}
\scriptsize
Accuracy, violation, and cost are conditional on power flow (PF) convergence. On the 300-bus system, PF convergence rates are 37.80\% for GNN, 98.50\% for MU, 99.30\% for QC, and 3.30\% for AS; all other entries have 100.00\% coverage.
\end{minipage}
\end{table*}
\subsection{Cross-System Performance}
\label{sec:crosssystem}
\begin{figure}[H]
    \centering
\includegraphics[width=1.0\linewidth]{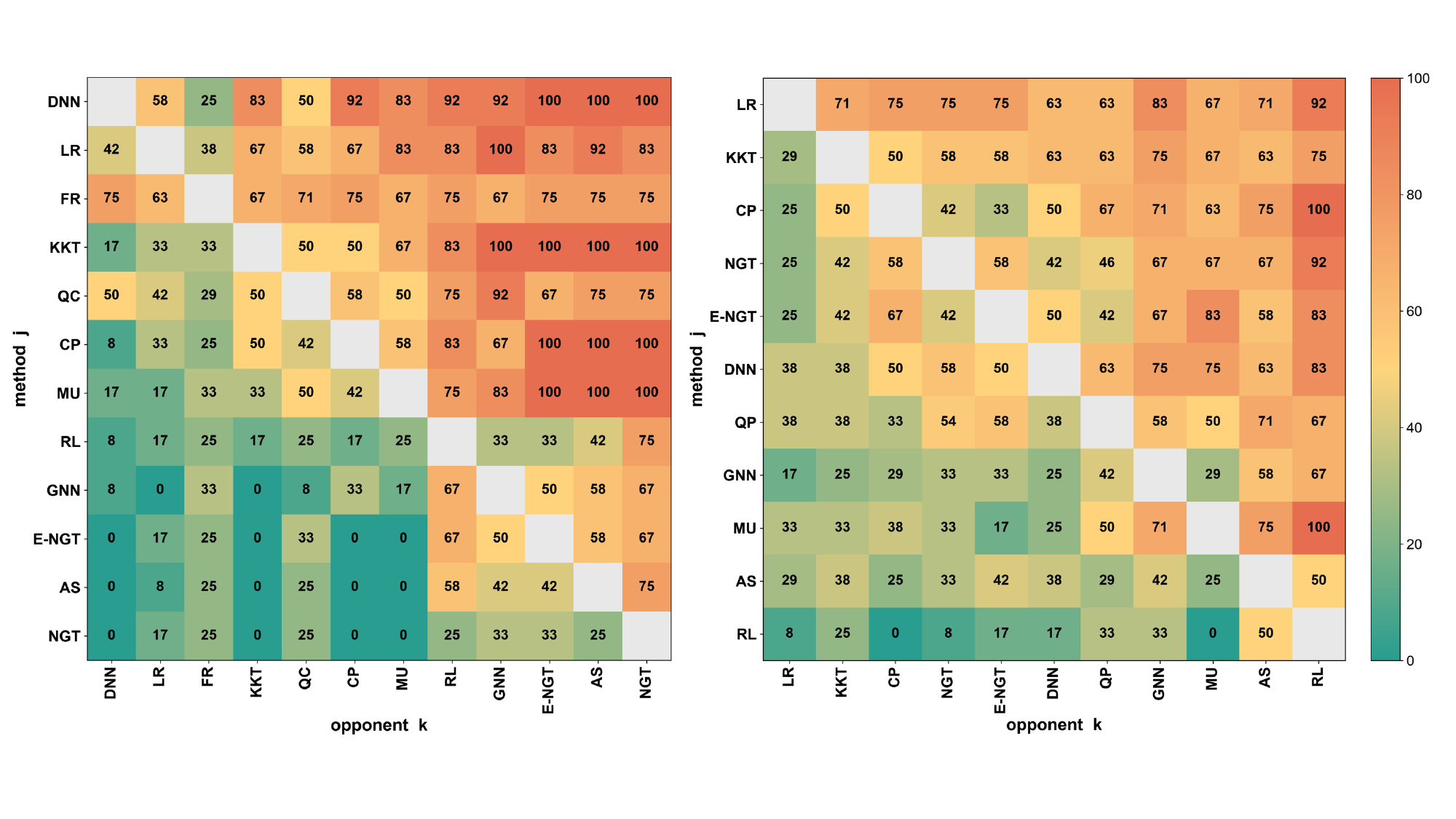}
    \caption{Pairwise win-rate matrices $W_{jk}$ for AC-OPF (left) and
DC-OPF (right). Each cell gives the win rate (\%) of the row method
$j$ against the column method $k$ over $U=12$ comparison units, where
one unit is one (system, objective) pair formed by the three test
systems (30/118/300-bus) and the four ranking objectives
($\mathrm{MAE}_{P_g}$, $\mathrm{Viol}_{\mathrm{total}}$,
$|\mathrm{Cost}_{\mathrm{gap}}|$, inference time).}
    \label{fig:winrate}
\end{figure}
Tables~\ref{tab:dc_stats} and~\ref{tab:ac_stats} report prediction accuracy,
composite constraint violation, cost gap, and inference time on the three test
systems under in-distribution conditions, and Table~\ref{tab:opf_rank}
condenses them into per-system Pareto levels and marginal win rates.
Per-component violations are given in Appendix~\ref{app:tables}.

Accuracy and cost do not certify feasibility. On the 300-bus DC system,
\texttt{LR} attains a cost gap of only $-0.003\%$, yet its prediction error
and composite violation are among the larger values in the table. Total cost
aggregates the dispatch into a single scalar, so a misallocation of generation can leave the cost almost unchanged while redistributing line flows. Constraint classes can also trade off against each other, as on the 118-bus AC
system, where \texttt{CP} lowers the reactive-power violation of \texttt{DNN}
but raises its voltage violation.

Larger test systems expose greater constraint violations. For \texttt{DNN}, they rise from near zero on 30-bus to their largest values on the 300-bus
system in both formulations. Over the same range, we found that the number of
active-set classes observed in training increases from 3 to 64 and 1032 in DC,
and from 2 to 5 and 2372 in AC (only the Pg limits), so the classification task becomes considerably more
diverse on the larger systems.

Feasibility restoration is bought with online computation. \texttt{QP}
projects the \texttt{CP} prediction onto the DC feasible region and reduces
violations to numerical precision on all three systems, at the price of an inference time that grows with system size. In AC,
\texttt{FR} follows the \texttt{CP} prediction with a solver-based OPF
recovery, which every test sample triggers. It closely matches the reference
solution on the 30- and 118-bus systems, but its inference time is one to two
orders of magnitude above that of \texttt{CP}. On the 300-bus system, recovery
fails on 108 of the 1000 test samples (10.80\%), yet \texttt{FR} still reduces
the violation of \texttt{CP} substantially.

For DC-OPF, \texttt{LR} is the only method on the first Pareto front of all
three systems, and four physics-informed methods (\texttt{KKT}, \texttt{CP},
\texttt{NGT}, and \texttt{E-NGT}) follow ahead of \texttt{DNN}. For AC-OPF,
\texttt{DNN}, \texttt{LR}, and \texttt{FR} share the first front on all three
systems; perhaps surprisingly, \texttt{LR} remains competitive despite the
nonconvexity of AC-OPF. Overall, no single method leads on every objective.
The first front is formed by methods that trade accuracy, feasibility, cost,
and speed against one another. Figure~\ref{fig:winrate} complements the aggregate rankings with
pairwise comparisons. Each entry reports the fraction of
system--metric comparisons in which the row method outperforms
the column method, with ties counted as half a win. Whether these in-distribution standings
survive a shift in the load distribution is examined next.

\begin{table}[h]
\centering
\caption{In-distribution ranking.}
\label{tab:opf_rank}
\scriptsize
\setlength{\tabcolsep}{3pt}
\begin{tabular}{llrrrrr}
\toprule
Rank & Method & 30-bus & 118-bus & 300-bus
& Mean & Marginal (\%) \\
\midrule
\multicolumn{7}{l}{DC-OPF} \\
\midrule
1  & LR    & 1 & 1 & 1 & 1.00 & 73.33 \\
2  & KKT   & 2 & 2 & 1 & 1.67 & 60.00 \\
3  & CP    & 3 & 1 & 1 & 1.67 & 57.50 \\
4  & NGT   & 3 & 1 & 1 & 1.67 & 56.25 \\
5  & E-NGT & 2 & 2 & 1 & 1.67 & 55.83 \\
6  & DNN   & 3 & 1 & 2 & 2.00 & 59.17 \\
7  & QP    & 4 & 1 & 1 & 2.00 & 50.42 \\
8  & GNN   & 3 & 1 & 2 & 2.00 & 35.83 \\
9  & MU    & 4 & 1 & 2 & 2.33 & 47.50 \\
10 & AS    & 2 & 2 & 3 & 2.33 & 35.00 \\
11 & RL    & 5 & 2 & 3 & 3.33 & 19.17 \\
\midrule
\multicolumn{7}{l}{AC-OPF} \\
\midrule
1  & DNN       & 1 & 1 & 1 & 1.00 & 79.55 \\
2  & LR        & 1 & 1 & 1 & 1.00 & 72.35 \\
3  & FR        & 1 & 1 & 1 & 1.00 & 71.21 \\
4  & KKT       & 2 & 1 & 1 & 1.33 & 66.67 \\
5  & QC  & 2 & 1 & 1 & 1.33 & 60.23 \\
6  & CP        & 2 & 2 & 1 & 1.67 & 60.61 \\
7  & MU  & 2 & 2 & 1 & 1.67 & 59.09 \\
8  & RL        & 3 & 2 & 1 & 2.00 & 28.79 \\
9  & GNN& 3 & 2 & 2 & 2.33 & 31.06 \\
10 & E-NGT     & 3 & 3 & 2 & 2.67 & 28.79 \\
11 & AS  & 3 & 3 & 3 & 3.00 & 25.00 \\
12 & NGT       & 4 & 3 & 3 & 3.33 & 16.67 \\
\bottomrule
\end{tabular}
\par\smallskip
\begin{minipage}{\columnwidth}
\scriptsize
System columns report Pareto levels for four objectives: prediction MAE, absolute cost gap,
composite violation, and inference time. Mean averages the three Pareto levels, and Marginal averages
pairwise win rates across competing methods. Methods are ordered by increasing Mean and then decreasing
Marginal.
\end{minipage}
\end{table}

\subsection{Load-Distribution Shifts}
\label{sec:generalization}

\begin{figure}[t]
  \centering
  \begin{subfigure}{\linewidth}
    \centering
    \includegraphics[width=\linewidth]{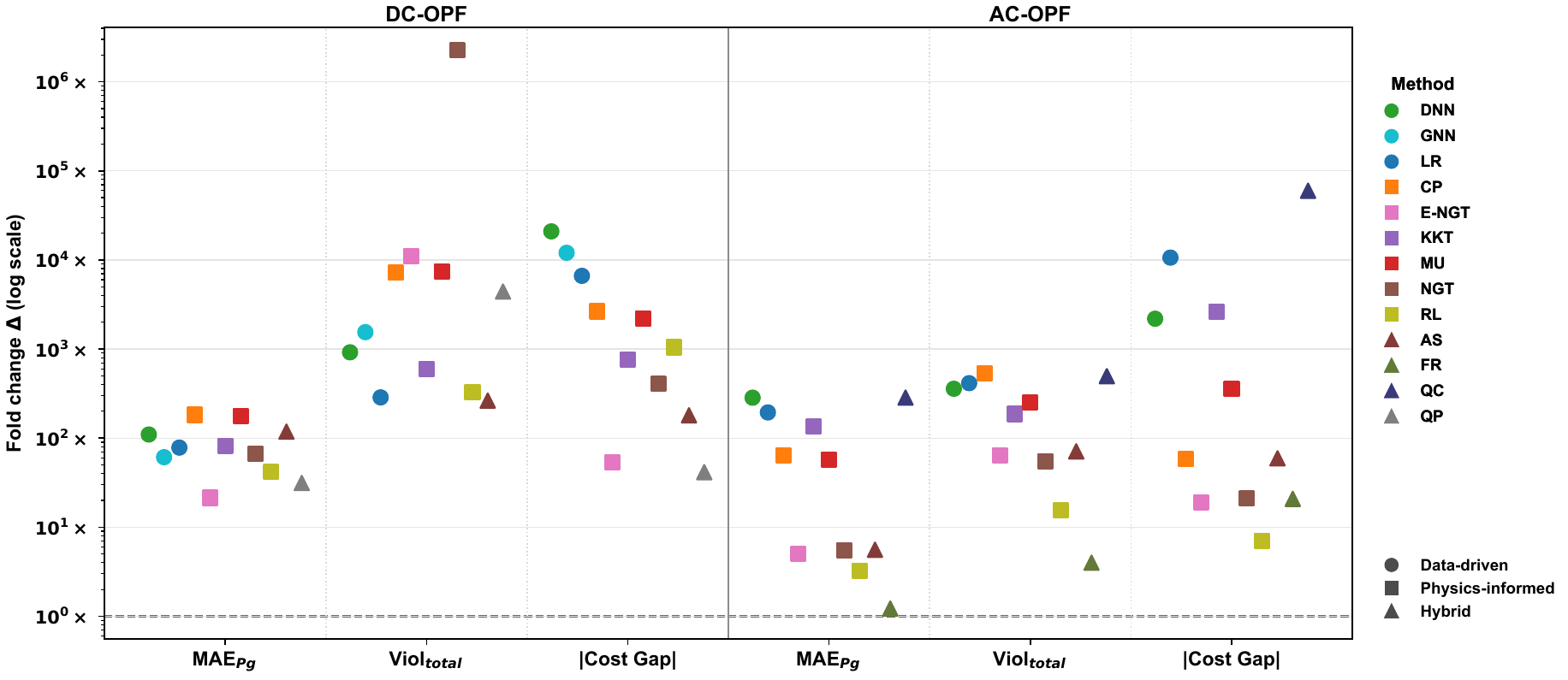}
    \caption{Heavier-load (API) shift.}
    \label{fig:generalization_heavier}
  \end{subfigure}

  \vspace{2pt}

  \begin{subfigure}{\linewidth}
    \centering
    \includegraphics[width=\linewidth]{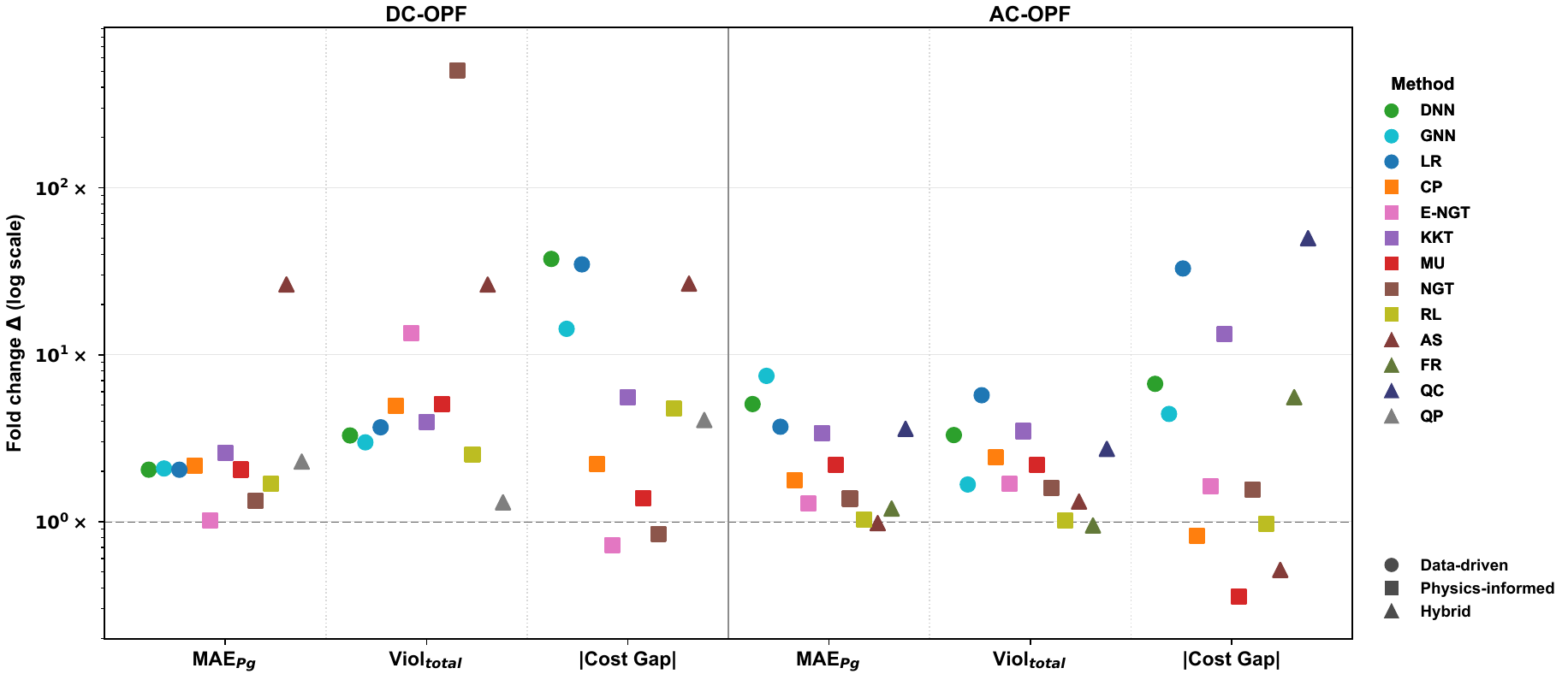}
    \caption{Larger-variance shift.}
    \label{fig:generalization_larger}
  \end{subfigure}

  \caption{Generalization degradation on the 118-bus system under two
  test-time load-distribution shifts. Each marker is the degradation factor
  $\Delta = E_{\mathrm{shift}}/E_{\mathrm{typ}}$ and the vertical axis is
  logarithmic. Marker colour identifies the method and marker shape encodes
  the ML paradigm.}
  \label{fig:generalization}
\end{figure}

We stress-test every method on the 118-bus system under two test-time
distribution shifts, a larger load variance and the Active Power Increase (API)
scenario, holding the trained model fixed.
Figures~\ref{fig:generalization_heavier} and~\ref{fig:generalization_larger}
report the ratio of each metric under the shift to its in-distribution value,
for prediction error, composite violation, and absolute cost gap.

Under the larger-variance shift, performance changes moderately: the
median ratio is about $3\times$ in DC-OPF and $2\times$ in AC-OPF, and every
ratio is smaller than its API counterpart. The API scenario causes a far more severe failure. All methods other than
\texttt{QP} and \texttt{FR} exhibit substantial violations; \texttt{DNN}, for
example, moves from 0.711\% to 78.020\% prediction error in DC, and
\texttt{GNN} fails to converge in AC power flow on every API sample. These
methods also report large negative cost gaps, which accompany infeasible
operating states rather than better solutions. The physics-informed
losses of \texttt{CP}, \texttt{MU}, \texttt{KKT}, \texttt{NGT}, and
\texttt{E-NGT} encode the constraints of the training scenario but do not
prevent this failure: their composite violations under the API shift are of
the same magnitude as that of \texttt{DNN}.

\texttt{QP} and \texttt{FR} are the only methods that
maintain low constraint violations under the API shift. \texttt{QP} projects onto the DC feasible region defined by the API generator
bounds, line-flow limits, and sample-specific power balance, using the API PTDF
matrix, while \texttt{FR} checks feasibility and performs OPF recovery using
the API constraints. Their advantage therefore includes the ability of online
optimization to adapt to new constraints and cannot be attributed entirely to
the generalization of the underlying predictor. As in
Section~\ref{sec:crosssystem}, this robustness is paid for with online
computation.

The overall generalization results imply that the benchmarked models largely
learn statistical correlations from the training distribution rather than the
underlying power-flow physics. The results also suggest that deployment-time
stress testing under congested, heavy-load regimes is essential before these
methods can be trusted in operation.

\begin{figure*}[t]
  \centering
  \begin{subfigure}{\linewidth}
    \centering
    \includegraphics[width=\linewidth]{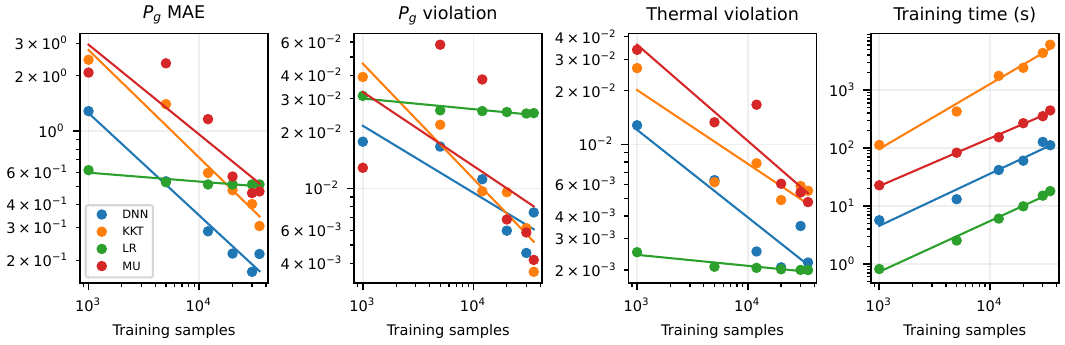}
    \caption{AC-OPF.}
    \label{fig:scaling_ac}
  \end{subfigure}

  \vspace{2pt}

  \begin{subfigure}{\linewidth}
    \centering
    \includegraphics[width=\linewidth]{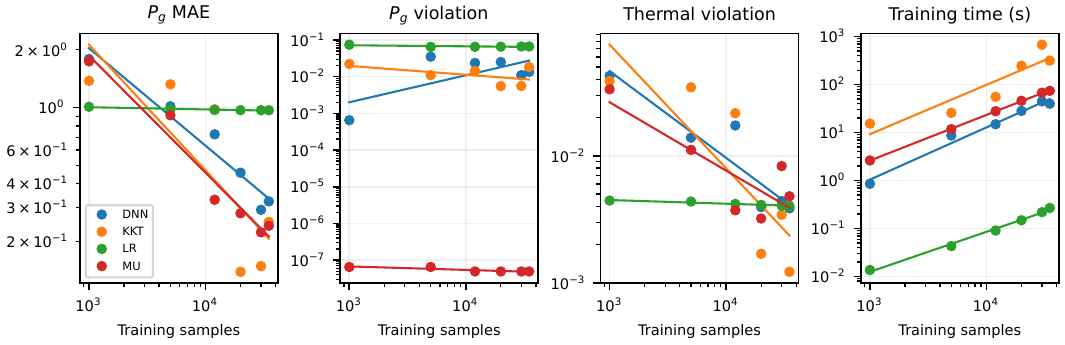}
    \caption{DC-OPF.}
    \label{fig:scaling_dc}
  \end{subfigure}

  \caption{Data-scaling behaviour on the 118-bus system as the training-set
  size grows.}
  \label{fig:scaling}
\end{figure*}

\subsection{Data Scaling Behavior}
\label{sec:scaling}
After examining performance across systems and load distributions, we next investigate how the amount of training data affects accuracy, feasibility, and
training cost on the 118-bus system. Figure~\ref{fig:scaling} shows prediction
error, active-power violation, line-flow violation, and training time for
training sets of 1K to 35K samples;
Appendix~\ref{app:data-scaling} gives the fitted parameters and the remaining AC
components.

With increasing training size, prediction error exhibits an overall decline trend for all neural-network-based methods evaluated, whereas \texttt{LR} changes little
($|\alpha| \leq 0.05$). \texttt{LR} achieves the lowest prediction error at 1K
samples in both formulations, but the neural methods overtake it as the
training set grows. \texttt{LR} is thus a competitive choice for prediction
accuracy when training data are limited, whereas the neural methods benefit
more from additional data over the examined range.

Feasibility and accuracy follow divergent trajectories. Line-flow violations
shrink with more data for most neural methods, but active-power violations in
DC do not decrease consistently. In AC, reactive-power violations decrease more
slowly than prediction error, and at 35K samples reactive power remains the
largest violation component for every tested method. More labeled data can
thus sharpen the predicted dispatch without a matching reduction in constraint
violations, particularly for AC reactive power.

Training time grows in proportion to the training set, with fitted
exponents between 0.82 and 1.12, and also differs widely between methods. At
35K samples, \texttt{KKT} trains 54 times longer than \texttt{DNN} in AC and is
still less accurate, whereas in DC its eightfold cost comes with about 22\% lower prediction error.

Validation-loss histories of the scaling runs are provided in Appendix~\ref{app:data-scaling}. Within each method, the lowest logged validation loss generally decreases with the training size, with smaller gains beyond about 20K samples for most methods, whereas the early-stopping epoch does not increase systematically with the data size.

\subsection{Compute-Budget Scaling Behavior}
\label{sec:compute}

\begin{figure}[t]
  \centering
  \begin{subfigure}{\linewidth}
    \centering
    \includegraphics[width=\linewidth]{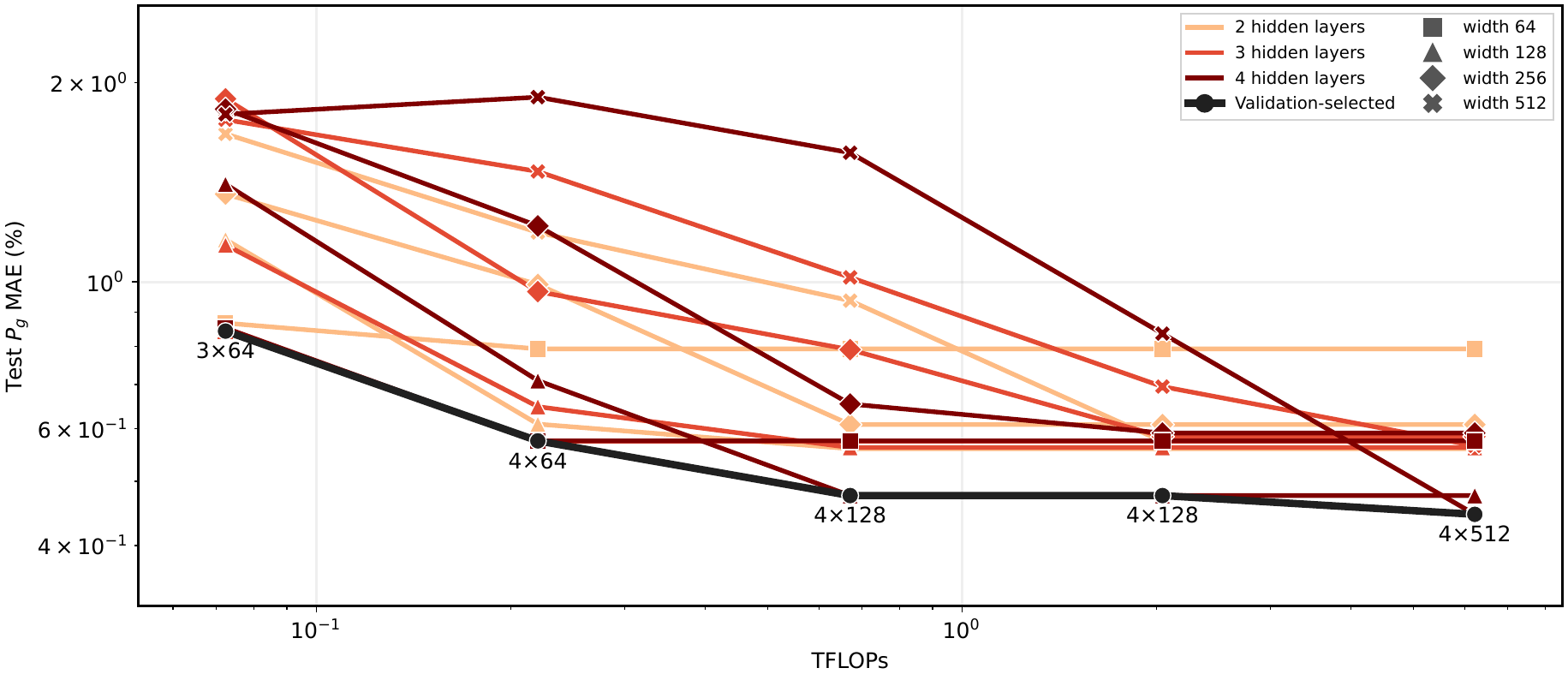}
    \caption{DC-OPF.}
    \label{fig:compute_budget_dc}
  \end{subfigure}

  \vspace{2pt}

  \begin{subfigure}{\linewidth}
    \centering
    \includegraphics[width=\linewidth]{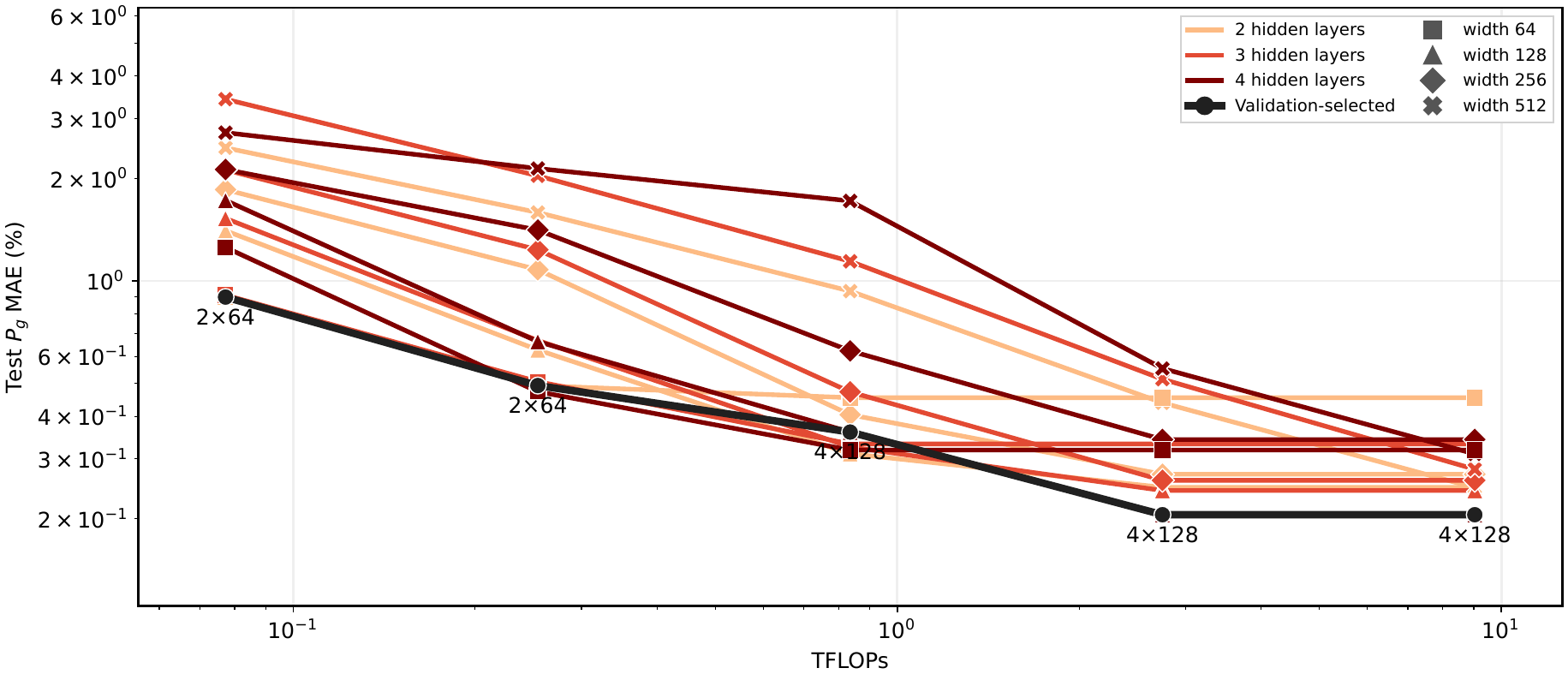}
    \caption{AC-OPF.}
    \label{fig:compute_budget_ac}
  \end{subfigure}

  \caption{Prediction accuracy under compute budgets on the
  118-bus system.
  Each coloured curve reports the test error of an architecture's
  best validation checkpoint available within each budget.
  Colour identifies hidden depth, and marker shape identifies width.
  The black curve reports the architecture selected by validation
  loss at each budget; annotations indicate depth $\times$ width.}
  \label{fig:compute-budget}
\end{figure}

We then fix the training set at 12,000 samples and vary the capacity of the
\texttt{DNN} instead, with hidden depths of 2, 3, and 4 and widths of 64 to 512,
giving 12 architectures per formulation. Five compute budgets, spaced
geometrically between the compute of one training epoch and that of the
complete run of the largest ($4\times512$) network, are fixed before test
evaluation. At each budget, the architecture with the lowest validation loss is
selected, and its test error is reported in Figure~\ref{fig:compute-budget}.
In the figure, the thick black curve follows this selection, and a curve turns
flat once the network has stopped early and can no longer use additional
budget.

At small budgets, the selected networks are small, since larger networks cannot
complete enough training within the budget; as the budget grows, the selection
moves to deeper and wider networks. Differences between architectures are
largest when compute is scarce and narrow as the budget grows. The architecture
with the lowest validation loss does not always achieve the lowest test
active-power error. At the intermediate AC budget of $0.835$ TFLOPs, the
selected $4\times128$ network has a test error of $0.360\%$, compared with
$0.311\%$ for the $2\times128$ network.

The scaling path shows diminishing returns. The error reduction per
additional TFLOP falls from 2.298
to 0.149 percentage points in AC and from 2.945 to 0.005 in DC. At larger
budgets, both formulations select four-hidden-layer networks, but widening
them helps little. In AC, the selected width-128 network outperforms the final
validation-selected checkpoint of the four-layer, width-512 network (0.206\%
versus 0.310\%), which uses nearly five times the compute; in DC, widening
from 128 to 512 lowers the error by only 6\% at about 15 times the compute.

In AC, the validation-selected network
trained on 12K samples achieves a slightly lower error than the data-scaling
\texttt{DNN} trained on 35K samples, whereas in DC the
35K-sample \texttt{DNN} remains more accurate (0.324\% versus 0.446\%). These
comparisons suggest that architecture selection and additional training data
offer different gains in the two formulations. Over the tested range, additional training data and larger compute budgets
generally improve prediction accuracy with diminishing returns, whereas
enlarging the network itself does not consistently help. More resources are
therefore beneficial, but spending additional compute solely on larger networks
is not always the most effective choice.

\section{Conclusion and Discussion}
\label{sec:conclusion}

We present ML-OPF-Bench, a unified and reproducible benchmark that evaluates
representative ML methods for AC- and DC-OPF under a consistent pipeline, a
scenario-based protocol spanning system sizes, distribution shifts, and data
and compute budgets, and a multi-objective ranking. The results show that in-distribution evaluation alone is an unreliable proxy
for deployable performance. Small errors and cost gaps can coexist with
constraint violations, and strong in-distribution performance does not ensure
robustness under heavily loaded, congested conditions. Additional data and
larger compute budgets improve accuracy with diminishing returns, whereas
larger networks bring uneven gains, and additional data alone does not secure
AC feasibility. 

In  the future
work, we plan to study how data quality, operating-condition coverage, and
network topology changes such as N-1 contingencies affect scaling across model
families. Building on these studies, we would like to explore scaling laws that
reflect the mathematical structure and active constraints of OPF itself rather than
relying on the generic power law, and to test them on further datasets and
power systems. We hope ML-OPF-Bench serves as an extensible, living platform
for integrating new algorithms, datasets, and evaluation scenarios toward
reliable, deployable ML-OPF tools.

\bibliographystyle{plainnat}
\bibliography{reference}

\clearpage
\onecolumn
\raggedbottom
\beginappendix

\section{Detailed Algorithm Implementations}
\label{app:methods}

This subsection describes the training objectives, model
representations, and recovery procedures used in the
benchmark.
The families are distinguished by their training
objectives and constraint-handling procedures.

\subsection{Pure Data-Driven Scheme}
The pure data-driven category refers to the absence of
explicit constraint penalties in the regression loss.
\begin{itemize}

\item \textbf{Linear Regression (LR).}
\texttt{LR} fits a multi-output linear least-squares
map from the load vector to the prediction
targets~\cite{linear}:
\begin{equation}
    \hat{\mathbf y}
    = \mathbf W\mathbf x+\mathbf b.
\end{equation}
The targets are non-slack active generation for DC,
and non-slack active generation together with
generator-bus voltage magnitudes for AC.

\item \textbf{Deep Neural Network (DNN).}
\texttt{DNN} uses a feedforward network trained with
MSE between predicted and reference setpoints,
denoted by $\mathcal L_{\mathrm{MSE}}$.
It serves as the baseline neural predictor.

\item \textbf{Graph Neural Network (GNN).}
The grid is represented as an undirected weighted graph.
The network applies Chebyshev spectral convolutions
with filters shared across nodes~\cite{GNN}:
\begin{equation}
  \mathbf{X}_{\ell} = \sigma_{\ell}\left(
    \sum_{k=0}^{K-1}
    \mathbf{T}_{k}\,\mathbf{X}_{\ell-1}\,\mathbf{W}_{\ell,k}
  \right),\quad \ell = 1,\dots,L.
\end{equation}
Here, $\mathbf{X}_{\ell}\in\mathbb{R}^{N_b\times F_{\ell}}$
is the node feature matrix at layer $\ell$;
$\mathbf{T}_{k}$ is the $k$-th Chebyshev polynomial of the
scaled graph shift operator, shared across all layers;
$\mathbf{W}_{\ell,k}\in\mathbb{R}^{F_{\ell-1}\times F_{\ell}}$
is the learnable filter coefficient matrix;
$K$ terms reach at most $K-1$ hops in each layer's receptive
field; $L$ is the number of graph-convolution layers; and
$\sigma_{\ell}$ is the ReLU activation.

For AC, edge weights follow a Gaussian kernel on branch
impedance~\cite{GNN}.
A DC-OPF dispatch is passed through AC power flow to
construct node features
$(V_i^{\mathrm{sub}},\theta_i^{\mathrm{sub}},
P_i^{\mathrm{sub}},Q_i^{\mathrm{sub}})$,
where $P_i^{\mathrm{sub}}$ and $Q_i^{\mathrm{sub}}$
are net bus injections.
The network predicts non-slack active generation.
Generator voltage-magnitude setpoints for the final
power-flow evaluation are retained from this preliminary
state.
For DC, the graph uses normalized branch-susceptance
magnitudes and the bus-level features described in the
benchmarking pipeline.

\end{itemize}

\subsection{Physics-Informed Scheme}

These methods incorporate physical constraints into
training losses or rewards.

\begin{itemize}

\item \textbf{Constraint Penalty (CP).}
\texttt{CP} augments supervised regression with
constraint-related penalty terms:
\begin{equation}
    \mathcal L_{\mathrm{CP}}
    =
    w_1\mathcal L_{\mathrm{MSE}}
    +w_2\mathcal L_{\mathrm{pen}}.
\end{equation}
For DC, the implemented branch term is
\begin{equation}
    \mathcal L_{\mathrm{pen}}^{\mathrm{DC}}
    =
    \mathbb E\left[
        \frac{1}{|\mathcal E_c|}
        \sum_{\ell\in\mathcal E_c}
        \left(
            \frac{f_\ell^2}{\bar f_\ell^2}-1
        \right)
    \right],
\end{equation}
where $\mathcal E_c$ contains the constrained branches
and $\bar f_\ell$ is the corresponding flow limit.

For AC, the penalty aggregates slack active- and
reactive-generation violations, PV reactive-generation
violations, PQ voltage violations, and branch thermal
violations.
Its gradient through the power-flow calculation is
estimated using a two-point zeroth-order finite
difference~\cite{DeepDCOPF,DeepACOPF}.

\item \textbf{KKT Embedded (KKT).}
\texttt{KKT} trains the network against the KKT optimality
conditions following~\cite{KKTAC,KKTDC}.
For AC, three network branches predict active and reactive
generation at all generators;
rectangular voltage components at all buses; and dual variables,
respectively. For evaluation, the predicted rectangular voltages are
converted to magnitudes and angles algebraically,
without a power-flow solve.
The supervised-state and residual terms are:
\begin{equation}
\begin{aligned}
  \mathcal{L}_{\mathrm{KKT}} ={}&
   \frac{1}{N_t}\sum_{i=1}^{N_t}\Big[
      \Lambda_P
      \|\hat{P}_{g,i}-P_{g,i},\,\hat{Q}_{g,i}-Q_{g,i}\| \\
    &\;+ \Lambda_V
      \|\hat{V}_{r,i}-V_{r,i},\,\hat{V}_{i,i}-V_{i,i}\|
      + \Lambda_\mu\|\hat{\mu}_i-\mu_i\| \Big] \\
    &\;+ \frac{\Lambda_\epsilon}{N_t+N_c}
       \sum_{i=1}^{N_t+N_c}
       \mathcal{L}_{\mathrm{KKT\text{-}res},i}.
\end{aligned}
\end{equation}
Here, $N_t$ is the number of supervised training data points
and $N_c$ is the number of unlabeled collocation points.
$\Lambda_P$ weights the combined active/reactive generation
prediction loss, $\Lambda_V$ the rectangular voltage prediction
loss ($V_r=V\cos\theta$, $V_i=V\sin\theta$), and $\Lambda_\mu$
the loss over exported dual variables $\mu$ (the power-balance
equality multipliers together with the generation, voltage,
and line-flow inequality multipliers).
The residual $\mathcal{L}_{\mathrm{KKT\text{-}res}}$%
~\cite{KKTAC,KKTDC} penalizes stationarity, complementary
slackness, and primal--dual feasibility.
DC predicts active powers and duals only.

\item \textbf{Multiplier Updates (MU).}
For AC, \texttt{MU} uses a shared input block followed by four
output heads predicting
$(\hat{\mathbf V}_m,\hat{\boldsymbol\theta},
\hat{\mathbf P}_g^{\mathrm{ns}},\hat{\mathbf Q}_g)$.
Voltage magnitudes and angles cover all buses, active generation
covers non-slack generators, and reactive generation covers
all generators. These quantities are predicted without a
power-flow solve during training.
\texttt{MU} casts OPF as a Lagrangian dual under
violation-based relaxation,
\begin{equation}
  \mathcal{L}_{\mathrm{MU}}
  = \mathcal{L}_{\mathrm{MSE}}
  + \sum_{c\in\mathcal{C}}\lambda_c\,\nu_c(\hat{y}),
\end{equation}
where $\lambda_c\geq 0$ is the Lagrange multiplier
(initialized to zero) and $\nu_c(\hat{y})$ is the violation
degree for constraint $c$.
The per-constraint multipliers are updated by a
subgradient step~\cite{lagrangian}.
AC violations cover voltage and angle bounds, generator limits,
thermal limits, Ohm's law deviations, and Kirchhoff's current
law balances, reducing to generator and line-flow limits for DC.

\item \textbf{NGT / E-NGT.} Our AC NGT and E-NGT implementations follow the original method, with stability adaptations for the benchmark data and training conditions. Following~\cite{unsupervised}, \texttt{NGT} learns without supervised solution targets,
using generation cost and physical residuals.
For AC, the network predicts voltages at non-zero-injection
buses, while Kron reduction and the nodal injection
equations reconstruct the remaining state.
The objective has the form:
  \begin{equation}
    \mathcal{L}_{\mathrm{NGT}} = k_{\text{obj}}\,\mathcal{L}_{\text{obj}}
      + \sum_{i=1}^{N_c} k_i\, \mathcal{L}_i 
  \end{equation}
with $\mathcal{L}_i = \mathbb{E}\big[\sum_j \nu_{ij}^2\big]$, where
$\nu_{ij}$ denotes the residual of the $j$-th constraint in category $i$,
equal to the violation amount beyond the bound for the inequality limits on
generation, voltage, and line flow, and to the mismatch of the
nodal power-balance equality. For AC, we modify the original design in response to the failure modes observed in our reproduction. First, evaluating the cost on the raw reconstructed $\hat{P}_g$ allows the
network to exploit out-of-bound dispatches with spuriously low cost, while
clamping $\hat{P}_g$ instead zeroes the cost gradient outside the generation
bounds. Second, the ratio-based weight update
  $k_i = \min(k_{\text{obj}}\mathcal{L}_{\text{obj}}/\mathcal{L}_i,\bar{k}_i)$
  saturates, locking $k_i$ at its upper bound once $\mathcal{L}_i$ vanishes or
  stalling it at the lower bound under persistent violations, and its loss
  magnitudes require per-case retuning. Accordingly, a linear marginal-cost
  extrapolation keeps a non-zero gradient that pushes $\hat{P}_g$ toward
  feasibility even when clamped, and a three-regime adaptive weighting (decay
  when satisfied, ratio formula in the normal range, proportional boost under
  severe violation) operates on EMA-smoothed losses with fixed normalization scales, using a common update rule across cases. 
  
  \texttt{E-NGT}, at each training epoch, is first updated on a small labeled set using a supervised voltage loss combined with
constraint penalties:
\begin{equation}
  \mathcal{L}_{\mathrm{E\text{-}NGT}} = k_v\,\mathcal{L}_v
    + \sum_{i=1}^{N_c} k_i\, \mathcal{L}_i 
\end{equation}
and then updated on all training samples under the
unsupervised objective $\mathcal{L}_{\text{NGT}}$. The coefficient $k_v$ is fixed, while the remaining weights are driven solely by the unsupervised step. In DC, physics loss uses raw generation cost and squared slack/branch violations with analogous adaptive weights. DC E-NGT supervises non-slack powers, replacing the AC voltage targets.

\item \textbf{Reinforcement Learning (RL).}
We implement a one-step DDPG baseline motivated by~\cite{rlopf}.
The policy is trained through interactions with the underlying grid
environment, without using labeled OPF solutions.
It controls non-slack $P_g$ within the physical generator bounds.
The reward is:
\begin{equation}
  r=-\tfrac12[1+\tanh(C/C_s)]-\frac{e}{1+e},
\end{equation}
where $C$ is the generation cost and $C_s$ is a cost scale derived
from physical generation bounds and cost coefficients.
The violation term $e$ sums component-mean violations
of active- and reactive-generation limits, voltage
bounds, thermal limits, and branch angle-difference
limits in AC.
DC omits the reactive-generation, voltage, and angle
terms and includes a power-balance violation term.
\end{itemize}

\subsection{Hybrid Scheme}

Hybrid methods combine learned predictions with
explicit post optimization or constraint-handling
procedures.

\begin{itemize}

\item \textbf{QP-Project (QP).}
For DC, \texttt{QP} pairs the \texttt{CP} predictor
with a projection onto the DC feasible
region~\cite{DeepDCOPF}:
\begin{equation}
    \mathbf P_g^\star
    =
    \underset{\mathbf P_g\in\mathcal F_{\mathrm{DC}}}
    {\operatorname{arg\,min}}
    \frac12\|\mathbf P_g-\hat{\mathbf P}_g\|_2^2.
\end{equation}
Here, $\hat{\mathbf P}_g$ includes the reconstructed
slack output, and $\mathcal F_{\mathrm{DC}}$
enforces generator bounds, power balance, and
PTDF-based branch-flow limits.
The convex quadratic problem is solved numerically
using SLSQP.

\item \textbf{Q-Correction (QC).}
For AC, the predictor is trained with MSE and uses
sigmoid outputs mapped to physical active-generation
and generator-voltage ranges~\cite{Qpost}.
After the initial power flow, reactive-limit
violations trigger a correction attempt.
A non-slack generator bus is converted from PV to
PQ only when all generators at that bus violate
their reactive-power limits.
Their reactive outputs are fixed at the corresponding
limits, and power flow is solved again.
The slack bus retains its original type.
Our implementation uses nominal voltage bounds
without tightening them during data generation.

\item \textbf{Feasibility Recovery (FR).}
For AC, \texttt{FR} augments the \texttt{CP}
prediction-and-power-flow pipeline with solver-based
OPF recovery~\cite{DeepACOPF}, using PYPOWER's
PIPS interior-point solver.
The screen checks power-flow convergence, generator
active and reactive limits, bus-voltage limits,
and both-end thermal limits. If the screen fails, an AC-OPF solve is attempted.
Failed recovery returns the original
power-flow result.

 \item \textbf{Active Set (AS).}
Following the active-set classification approach
in~\cite{Deka2019LearningFD}, a ReLU classifier is trained
with cross-entropy to predict active-set classes.
In our implementation, each class corresponds to a binary
vector constructed from thresholded dual variables:
\begin{equation}
    \mathbf{a}
    =
    \bigl(\mathbb{I}\{\mu_j>\epsilon\}\bigr)_{j\in\mathcal{C}},
\end{equation}
where $\mu_j$ is the multiplier of inequality constraint $j$.
For DC, $\mathcal{C}$ includes generator active-power bounds
and branch-flow limits; for AC, it includes only generator
active-power bounds. Up to three predicted classes are considered for recovery.
The original DC method reconstructs each candidate through
a linear system~\cite{Deka2019LearningFD}.
We use LP-based recovery to accommodate predicted active sets
that do not uniquely determine the dispatch.
For DC, the lowest-cost recovered candidate passing the
feasibility checks is retained.
For our AC extension, a simplified LP recovers non-slack
active-generation setpoints, while generator-voltage setpoints
are fixed at their training-set means. AC candidates are screened against non-slack active-generation bounds and ranked by their non-slack generation cost. The selected candidate is then evaluated through the common AC power-flow pipeline.
\end{itemize}

\clearpage
\section{Additional Data-Scaling Metrics and Training Histories}
\label{app:data-scaling}

This subsection supplements Section~\ref{sec:scaling} with
additional constraint-violation metrics, fitted scaling
parameters, and training histories.
These results provide a fuller view of how prediction
accuracy, constraint performance, and training cost
change with training-set size.

The data-scaling model is $m(N)=A(N/N_0)^\alpha$, with
$N_0=10{,}000$ samples as a reference scale.
The primary fit, used for the plotted curves, minimizes
$\sum_i[\ln m_i-\ln A-\alpha\ln(N_i/N_0)]^2$. $R^2_{\log}$ measures goodness of
fit using log-transformed metric values, whereas
$R^2_{\mathrm{raw}}$ uses the original metric values.

As a sensitivity check, we also fit the same model
by nonlinear least squares (NLS), minimizing
$\sum_i[m_i-A(N_i/N_0)^\alpha]^2$.
We report its exponent $\alpha_{\mathrm{NLS}}$ and
its goodness of fit on the original scale,
$R^2_{\mathrm{raw,NLS}}$.
Thus, $R^2_{\log}$ and $R^2_{\mathrm{raw}}$ evaluate
the primary curve on two scales, while
$R^2_{\mathrm{raw,NLS}}$ evaluates the separately
fitted NLS curve on the original scale.

All fits use each series' available observations,
and plotted curves remain within its observed
training-size interval.

\subsection*{Data Scaling}
\scriptsize
\setlength{\tabcolsep}{3pt}
\begin{longtable}{lllrrrrrr}
\toprule
Form & Method & Metric & $A$ & $\alpha$
& $R^2_{\log}$ & $R^2_{\mathrm{raw}}$
& $\alpha_{\mathrm{NLS}}$ & $R^2_{\mathrm{raw,NLS}}$ \\
\midrule
\endhead
AC & DNN & MAE $P_g$ (\%) & 0.349 & -0.555 & 0.974 & 0.995 & -0.568 & 0.996 \\
AC & KKT & MAE $P_g$ (\%) & 0.719 & -0.584 & 0.966 & 0.939 & -0.508 & 0.965 \\
AC & LR & MAE $P_g$ (\%) & 0.534 & -0.048 & 0.826 & 0.837 & -0.051 & 0.839 \\
AC & MU & MAE $P_g$ (\%) & 0.960 & -0.487 & 0.786 & 0.480 & -0.328 & 0.648 \\
AC & DNN & Viol $P_g$ & 0.009 & -0.356 & 0.742 & 0.692 & -0.280 & 0.754 \\
AC & KKT & Viol $P_g$ & 0.011 & -0.613 & 0.920 & 0.914 & -0.508 & 0.965 \\
AC & LR & Viol $P_g$ & 0.026 & -0.056 & 0.860 & 0.867 & -0.059 & 0.869 \\
AC & MU & Viol $P_g$ & 0.013 & -0.394 & 0.245 & -0.124 & -0.154 & 0.073 \\
AC & DNN & Viol line & 0.004 & -0.488 & 0.837 & 0.950 & -0.508 & 0.958 \\
AC & KKT & Viol line & 0.008 & -0.412 & 0.774 & 0.823 & -0.558 & 0.918 \\
AC & LR & Viol line & 0.002 & -0.061 & 0.872 & 0.879 & -0.064 & 0.881 \\
AC & MU & Viol line & 0.010 & -0.537 & 0.870 & 0.899 & -0.478 & 0.913 \\
AC & DNN & Train (s) & 35.96 & 0.898 & 0.962 & 0.919 & 1.053 & 0.943 \\
AC & KKT & Train (s) & 1267.40 & 1.119 & 0.984 & 0.963 & 1.367 & 0.984 \\
AC & LR & Train (s) & 5.50 & 0.875 & 0.992 & 0.986 & 1.021 & 0.999 \\
AC & MU & Train (s) & 146.93 & 0.824 & 0.997 & 0.989 & 0.905 & 0.993 \\
DC & DNN & MAE $P_g$ (\%) & 0.633 & -0.508 & 0.947 & 0.938 & -0.436 & 0.973 \\
DC & KKT & MAE $P_g$ (\%) & 0.471 & -0.655 & 0.653 & 0.284 & -0.372 & 0.639 \\
DC & LR & MAE $P_g$ (\%) & 0.977 & -0.011 & 0.910 & 0.911 & -0.012 & 0.911 \\
DC & MU & MAE $P_g$ (\%) & 0.457 & -0.611 & 0.959 & 0.963 & -0.557 & 0.971 \\
DC & DNN & Viol $P_g$ & 0.011 & 0.735 & 0.461 & -0.833 & 0.088 & 0.043 \\
DC & KKT & Viol $P_g$ & 0.011 & -0.240 & 0.296 & 0.357 & -0.230 & 0.385 \\
DC & LR & Viol $P_g$ & 0.067 & -0.028 & 0.548 & 0.572 & -0.030 & 0.574 \\
DC & MU & Viol $P_g$ & 0.000 & -0.095 & 0.781 & 0.769 & -0.093 & 0.770 \\
DC & DNN & Viol line & 0.010 & -0.683 & 0.857 & 0.911 & -0.593 & 0.931 \\
DC & KKT & Viol line & 0.008 & -0.974 & 0.699 & -0.293 & -0.449 & 0.698 \\
DC & LR & Viol line & 0.004 & -0.029 & 0.941 & 0.938 & -0.029 & 0.938 \\
DC & MU & Viol line & 0.008 & -0.538 & 0.689 & 0.878 & -0.653 & 0.947 \\
DC & DNN & Train (s) & 12.78 & 1.088 & 0.979 & 0.926 & 0.872 & 0.959 \\
DC & KKT & Train (s) & 98.63 & 1.027 & 0.832 & 0.547 & 1.183 & 0.662 \\
DC & LR & Train (s) & 0.09 & 0.833 & 0.994 & 0.983 & 0.984 & 0.997 \\
DC & MU & Train (s) & 23.30 & 0.951 & 1.000 & 0.998 & 0.923 & 0.998 \\
AC & DNN & MAE $V$ (\%) & 0.045 & -0.405 & 0.965 & 0.986 & -0.425 & 0.989 \\
AC & KKT & MAE $V$ (\%) & 0.115 & -0.265 & 0.976 & 0.984 & -0.253 & 0.987 \\
AC & LR & MAE $V$ (\%) & 0.070 & -0.048 & 0.828 & 0.838 & -0.050 & 0.839 \\
AC & MU & MAE $V$ (\%) & 0.068 & -0.382 & 0.965 & 0.943 & -0.339 & 0.963 \\
AC & DNN & Viol $Q_g$ & 0.067 & -0.154 & 0.514 & 0.546 & -0.149 & 0.549 \\
AC & KKT & Viol $Q_g$ & 0.106 & -0.234 & 0.877 & 0.926 & -0.252 & 0.933 \\
AC & LR & Viol $Q_g$ & 0.076 & -0.080 & 0.839 & 0.852 & -0.086 & 0.856 \\
AC & MU & Viol $Q_g$ & 0.044 & -0.361 & 0.902 & 0.847 & -0.300 & 0.893 \\
AC & DNN & Viol $V$ & 0.001 & -0.457 & 0.626 & 0.833 & -0.558 & 0.897 \\
AC & KKT & Viol $V$ & 0.001 & -0.290 & 0.743 & 0.857 & -0.304 & 0.863 \\
AC & LR & Viol $V$ & 0.001 & -0.100 & 0.818 & 0.837 & -0.110 & 0.845 \\
AC & MU & Viol $V$ & 0.001 & -0.390 & 0.551 & 0.697 & -0.562 & 0.841 \\
\bottomrule
\end{longtable}

\begin{figure}
    \centering
    \includegraphics[width=1.0\linewidth]{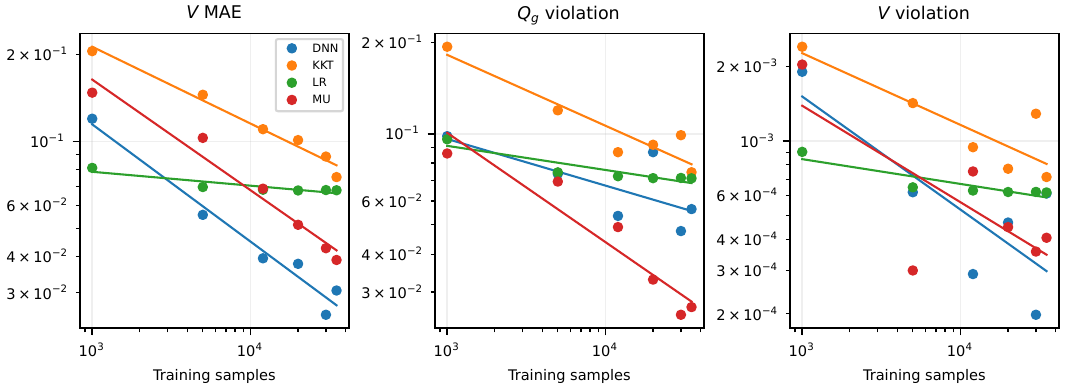}
    \caption{Additional AC scaling metrics: generator-voltage MAE, reactive-power violations, and voltage violations. Markers are observations; lines follow the common log-space fitting rule.}\label{fig:ac-supp}
\end{figure}

\begin{figure}
    \centering
    \includegraphics[width=1.0\linewidth]{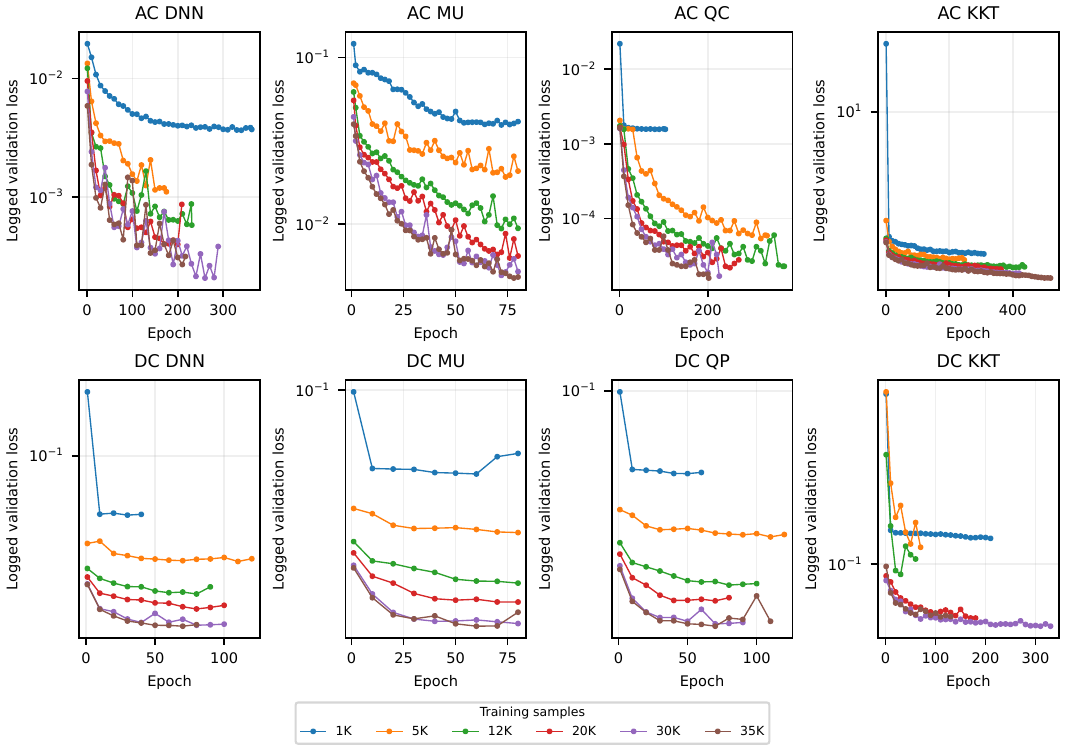}
   \caption{Validation histories for all neural scaling runs. Markers show printed epoch/loss pairs.}\label{fig:dynamics}
\end{figure}

\clearpage\section{Compute Protocol, Inventory, and Iso-error Maps}\label{app:compute}
\normalsize
This subsection supplements Section~\ref{sec:compute} with
the detailed training protocol, FLOP accounting,
experiment inventory, and iso-error maps.

The compute study fixes the 118-bus data and saved split arrays. AC uses 198 inputs and 107 outputs (53 non-slack powers and 54 generator voltages); DC uses 118 inputs and 53 non-slack powers. Each formulation uses 12,000 training samples; validation and test contain 5,000 samples each for AC and 4,953 each for DC. Hidden layers use ReLU. Training uses Adam with learning rate 0.001, batch size 64 including the final partial batch, and single-tensor updates. The cap is 1,000 epochs; every trajectory stops earlier under patience 20 and an absolute validation-improvement threshold of $10^{-6}$.

Let $d_0,\ldots,d_L$ include the input, all hidden layers, and the output, and let $P=\sum_{\ell=1}^{L}(d_{\ell-1}d_\ell+d_\ell)$. For a batch of $b$ samples, forward propagation including mean MSE costs,
\begin{equation}
F_f(b)=\sum_{\ell=1}^{L}(2b d_{\ell-1}d_\ell+b d_\ell)+3b d_L.
\end{equation}
Backward propagation, including mean-MSE derivatives, costs,
\begin{equation}
F_b(b)=\sum_{\ell=1}^{L}2b d_{\ell-1}d_\ell+\sum_{\ell=2}^{L}2b d_{\ell-1}d_\ell+\sum_{\ell=1}^{L}(b-1)d_\ell+2b d_L+2.
\end{equation}
The raw inputs require no gradient. Single-tensor Adam costs $F_A=13P+2L$ per update. An epoch sums $F_f(b)+F_b(b)+F_A$ over the actual batches, then adds $F_f(n_{\rm val})$ for validation. For 12,000 samples this includes 187 batches of 64 and one batch of 32. One multiply-accumulate counts as two FLOPs; square root and division each count as one reference arithmetic operation.

\subsection*{Compute-Study Trajectories}
\scriptsize

\begin{longtable}{lrrrrrrr}
\label{tab:compute-inventory}\\\toprule
Form & Depth & Width & Parameters & Stop epoch & Best epoch & TFLOPs & Train (s)\\\midrule\endhead
AC & 2 & 64 & 23851 & 214 & 194 & 0.365 & 58.41\\
AC & 2 & 128 & 55787 & 250 & 230 & 1.023 & 65.71\\
AC & 2 & 256 & 144235 & 171 & 151 & 1.870 & 47.60\\
AC & 2 & 512 & 419435 & 228 & 208 & 7.509 & 62.11\\
AC & 3 & 64 & 28011 & 333 & 313 & 0.684 & 107.31\\
AC & 3 & 128 & 72299 & 250 & 230 & 1.370 & 78.25\\
AC & 3 & 256 & 210027 & 161 & 141 & 2.653 & 49.60\\
AC & 3 & 512 & 682091 & 123 & 103 & 6.776 & 37.49\\
AC & 4 & 64 & 32171 & 333 & 313 & 0.800 & 122.76\\
AC & 4 & 128 & 88811 & 272 & 252 & 1.868 & 101.58\\
AC & 4 & 256 & 275819 & 131 & 111 & 2.885 & 42.98\\
AC & 4 & 512 & 944747 & 117 & 97 & 9.037 & 43.75\\
DC & 2 & 64 & 15221 & 121 & 101 & 0.133 & 28.32\\
DC & 2 & 128 & 38581 & 106 & 86 & 0.305 & 24.27\\
DC & 2 & 256 & 109877 & 83 & 63 & 0.707 & 19.21\\
DC & 2 & 512 & 350773 & 90 & 70 & 2.528 & 20.21\\
DC & 3 & 64 & 19381 & 107 & 87 & 0.155 & 27.96\\
DC & 3 & 128 & 55093 & 87 & 67 & 0.371 & 22.73\\
DC & 3 & 256 & 175669 & 68 & 48 & 0.956 & 18.21\\
DC & 3 & 512 & 613429 & 86 & 66 & 4.318 & 23.56\\
DC & 4 & 64 & 23541 & 91 & 71 & 0.163 & 27.13\\
DC & 4 & 128 & 71605 & 74 & 54 & 0.418 & 23.28\\
DC & 4 & 256 & 241461 & 70 & 50 & 1.371 & 21.70\\
DC & 4 & 512 & 876085 & 86 & 66 & 6.221 & 27.43\\
\bottomrule\end{longtable}

\begin{figure}
    \centering
    \includegraphics[width=1.0\linewidth]{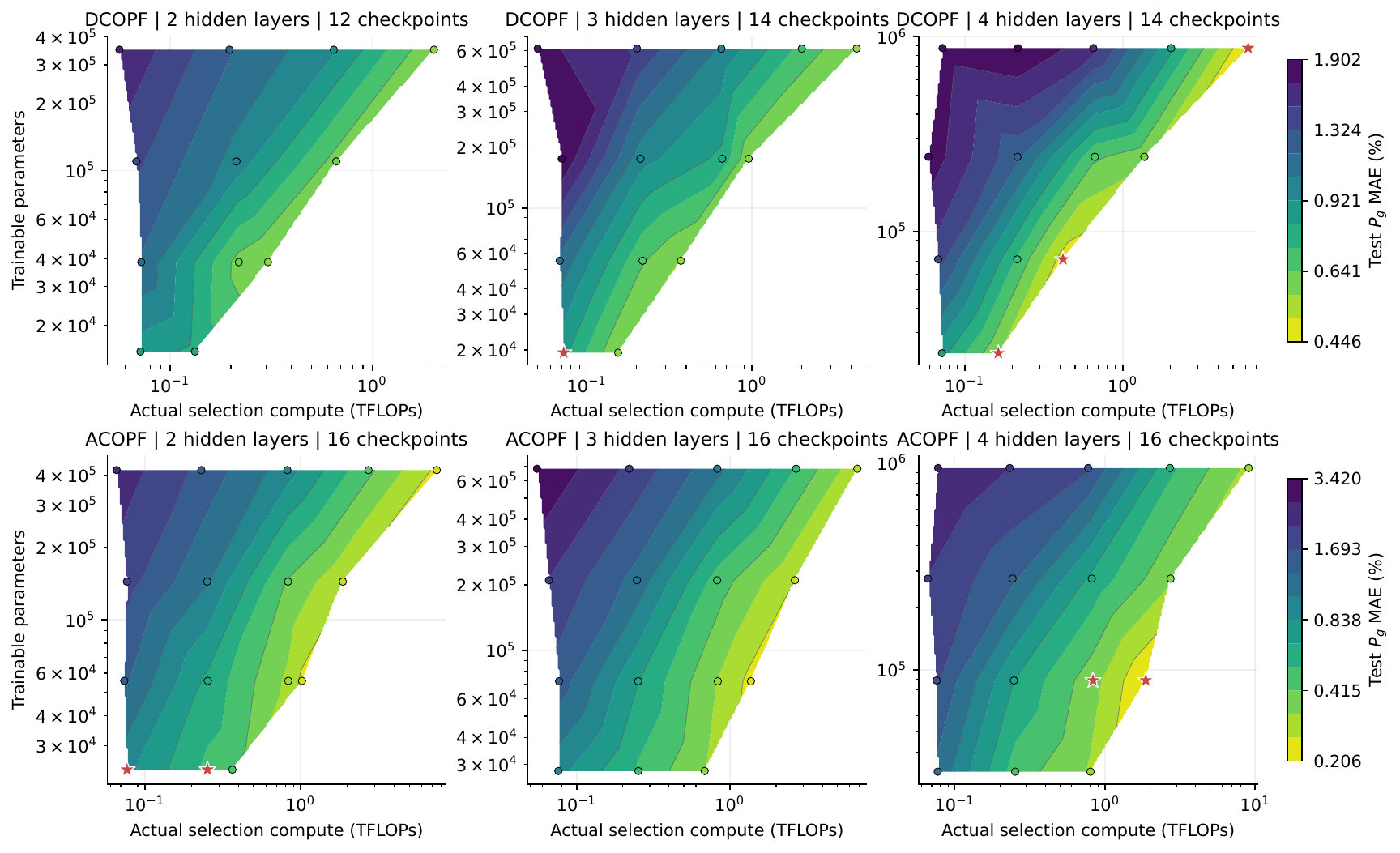}
    \caption{Iso-error maps for 118-bus DNNs, separated by formulation and hidden depth. Dots are distinct measured checkpoints; Red stars mark checkpoints selected by validation loss across
all architectures at the prescribed compute budgets. Colours show test $P_g$ MAE, interpolated between measured
checkpoints in log--log coordinates for compute $C$ and
parameter count $P$. Interpolation is linear in $\log_{10}$ MAE.}\label{fig:compute-contours}
\end{figure}

\subsection*{Validation-Selected Candidates}
\scriptsize
\begin{longtable}{lrrlrr}
\label{tab:compute-frontier}\\\toprule
Form & Budget (TFLOPs) & Actual (TFLOPs) & Depth $\times$ width & Checkpoint epoch & $P_g$ MAE (\%)\\\midrule\endhead
AC & 0.077 & 0.077 & $2\times 64$ & 45 & 0.896\\
AC & 0.254 & 0.252 & $2\times 64$ & 148 & 0.493\\
AC & 0.835 & 0.831 & $4\times 128$ & 121 & 0.360\\
AC & 2.748 & 1.868 & $4\times 128$ & 252 & 0.206\\
AC & 9.037 & 1.868 & $4\times 128$ & 252 & 0.206\\
DC & 0.072 & 0.072 & $3\times 64$ & 45 & 0.843\\
DC & 0.220 & 0.163 & $4\times 64$ & 71 & 0.576\\
DC & 0.671 & 0.418 & $4\times 128$ & 54 & 0.476\\
DC & 2.043 & 0.418 & $4\times 128$ & 54 & 0.476\\
DC & 6.221 & 6.221 & $4\times 512$ & 66 & 0.446\\
\bottomrule\end{longtable}
 
\clearpage\section{Complete Numerical Results}\label{app:tables}
\normalsize
This subsection provides the complete numerical results underlying the main-text figures and analyses, including cross-system comparisons, distribution shifts, and
scaling experiments. It also reports solution times of a conventional OPF
solver as a reference for computational performance. In the Scenario column, TYP denotes the typical in-distribution test condition
(relative standard deviation of the load perturbations 0.12), VAR the larger-variance
shift (relative standard deviation 0.25 with the original operating limits), and API
the heavily loaded Active Power Increase shift taken from the corresponding PGLib API
case with its changed loading conditions and operating limits; all models are trained
on TYP data only, and VAR and API are evaluated on the 118-bus system.

\subsection*{DC Cross-System and Scenarios}
\scriptsize
\begin{longtable}{lllrrrrrrrr}
\toprule
Case & Scenario & Method & MAE $P_g$  & Viol $P_g$ & Viol $P_g^0$ & Viol line & Gap (\%) & Infer (ms) & Train (s) & Epoch/steps \\
\midrule\endhead
30 & TYP & AS & 0.000 & 0.000 & 0.000 & 0.000 & 0.000 & 4.442 & 12.73 & 33 \\
30 & TYP & CP & 1.118 & 0.000 & 0.000 & 0.003 & -0.109 & 0.448 & 33.50 & 72 \\
30 & TYP & DNN & 0.162 & 0.000 & 0.000 & 0.000 & 0.040 & 0.454 & 14.53 & 51 \\
30 & TYP & E-NGT & 0.225 & 0.000 & 0.000 & 0.001 & -0.069 & 0.379 & 2732.66 & 2500 \\
30 & TYP & GNN & 1.369 & 0.000 & 0.000 & 0.003 & -0.034 & 2.950 & 164.73 & 118 \\
30 & TYP & KKT & 0.058 & 0.000 & 0.000 & 0.000 & -0.015 & 0.438 & 85.84 & 63 \\
30 & TYP & LR & 0.000 & 0.000 & 0.000 & 0.000 & 0.000 & 0.119 & 0.02 & -- \\
30 & TYP & MU & 1.134 & 0.000 & 0.000 & 0.003 & -0.079 & 0.463 & 41.99 & 80 \\
30 & TYP & NGT & 0.420 & 0.000 & 0.000 & 0.002 & -0.128 & 0.383 & 1809.24 & 2500 \\
30 & TYP & QP & 0.373 & 0.000 & 0.000 & 0.000 & 0.113 & 0.973 & 33.50 & 72 \\
30 & TYP & RL & 80.187 & 0.000 & 0.000 & 0.327 & -24.500 & 0.560 & 3386.54 & 760001 steps \\
118 & TYP & AS & 0.653 & 0.000 & 0.000 & 0.019 & -0.399 & 4.608 & 13.54 & 70 \\
118 & TYP & CP & 0.364 & 0.000 & 0.000 & 0.005 & 0.011 & 0.451 & 24.24 & 103 \\
118 & TYP & DNN & 0.711 & 0.021 & 0.000 & 0.009 & 0.001 & 0.439 & 13.36 & 94 \\
118 & TYP & E-NGT & 3.057 & 0.000 & 0.000 & 0.003 & 0.574 & 0.415 & 1644.82 & 3000 \\
118 & TYP & GNN & 1.073 & 0.005 & 0.000 & 0.016 & -0.003 & 2.906 & 152.53 & 214 \\
118 & TYP & KKT & 0.838 & 0.035 & 0.000 & 0.013 & -0.040 & 0.440 & 87.82 & 127 \\
118 & TYP & LR & 1.013 & 0.072 & 0.000 & 0.004 & -0.004 & 0.121 & 0.08 & -- \\
118 & TYP & MU & 0.377 & 0.000 & 0.000 & 0.004 & 0.014 & 0.483 & 22.17 & 80 \\
118 & TYP & NGT & 0.994 & 0.000 & 0.000 & 0.000 & 0.075 & 0.388 & 1108.16 & 3000 \\
118 & TYP & QP & 0.421 & 0.000 & 0.000 & 0.000 & 0.092 & 4.511 & 24.24 & 103 \\
118 & TYP & RL & 1.228 & 0.000 & 0.000 & 0.024 & -0.028 & 0.558 & 1531.15 & 300001 steps \\
118 & API  & AS & 77.203 & 0.000 & 0.000 & 4.950 & -71.908 & 4.315 & 13.54 & 70 \\
118 & API  & CP & 66.399 & 2.060 & 28.444 & 2.116 & -30.344 & 0.449 & 24.24 & 103 \\
118 & API  & DNN & 78.020 & 8.133 & 17.995 & 1.457 & -29.196 & 0.447 & 13.36 & 94 \\
118 & API  & E-NGT & 65.367 & 2.060 & 30.424 & 2.310 & -30.517 & 0.381 & 1644.82 & 3000 \\
118 & API  & GNN & 65.538 & 2.060 & 27.991 & 2.063 & -30.310 & 2.977 & 152.53 & 214 \\
118 & API  & KKT & 68.787 & 2.061 & 24.791 & 1.693 & -30.176 & 0.449 & 87.82 & 127 \\
118 & API  & LR & 79.277 & 7.678 & 12.695 & 1.455 & -28.923 & 0.120 & 0.08 & -- \\
118 & API & MU & 66.425 & 2.060 & 28.462 & 2.117 & -30.346 & 0.472 & 22.17 & 80 \\
118 & API & NGT & 66.150 & 2.060 & 27.881 & 2.098 & -30.304 & 0.408 & 1108.16 & 3000 \\
118 & API & QP & 13.174 & 0.000 & 0.000 & 0.000 & 3.805 & 7.350 & 24.24 & 103 \\
118 & API & RL & 51.063 & 0.000 & 7.106 & 0.758 & -28.949 & 0.540 & 1531.15 & 300001 steps \\
118 & VAR & AS & 17.218 & 0.000 & 0.000 & 0.494 & -10.639 & 4.894 & 13.54 & 70 \\
118 & VAR & CP & 0.791 & 0.000 & 0.000 & 0.022 & -0.025 & 0.454 & 24.24 & 103 \\
118 & VAR & DNN & 1.455 & 0.074 & 0.000 & 0.025 & -0.052 & 0.451 & 13.36 & 94 \\
118 &  VAR & E-NGT & 3.098 & 0.000 & 0.000 & 0.043 & 0.416 & 0.394 & 1644.82 & 3000 \\
118 &  VAR & GNN & 2.233 & 0.018 & 0.000 & 0.044 & -0.036 & 2.934 & 152.53 & 214 \\
118 & VAR & KKT & 2.159 & 0.138 & 0.000 & 0.053 & -0.220 & 0.446 & 87.82 & 127 \\
118 &  VAR & LR & 2.073 & 0.264 & 0.000 & 0.016 & -0.152 & 0.119 & 0.08 & -- \\
118 &  VAR & MU & 0.773 & 0.000 & 0.000 & 0.022 & -0.019 & 0.466 & 22.17 & 80 \\
118 &  VAR & NGT & 1.328 & 0.000 & 0.000 & 0.007 & 0.063 & 0.402 & 1108.16 & 3000 \\
118 &  VAR & QP & 0.962 & 0.000 & 0.000 & 0.000 & 0.373 & 5.132 & 24.24 & 103 \\
118 &  VAR & RL & 2.069 & 0.000 & 0.000 & 0.061 & -0.131 & 0.547 & 1531.15 & 300001 steps \\
300 & TYP & AS & 26.294 & 0.000 & 0.000 & 3.283 & 3.285 & 5.790 & 4.22 & 37 \\
300 & TYP & CP & 1.170 & 0.000 & 0.477 & 0.073 & 0.019 & 0.481 & 35.03 & 152 \\
300 & TYP & DNN & 1.789 & 0.274 & 0.618 & 0.099 & -0.152 & 0.456 & 10.43 & 140 \\
300 & TYP & E-NGT & 1.697 & 0.000 & 0.003 & 0.018 & 0.878 & 0.388 & 741.26 & 2500 \\
300 & TYP & GNN & 3.171 & 0.015 & 1.139 & 0.194 & 0.015 & 3.008 & 496.72 & 226 \\
300 & TYP & KKT & 1.540 & 0.182 & 0.685 & 0.100 & -0.071 & 0.447 & 83.18 & 234 \\
300 & TYP & LR & 2.413 & 0.444 & 0.000 & 0.046 & -0.003 & 0.128 & 0.16 & -- \\
300 & TYP & MU & 1.925 & 0.000 & 0.104 & 0.081 & 0.928 & 0.473 & 29.35 & 80 \\
300 & TYP & NGT & 4.586 & 0.000 & 0.001 & 0.017 & 0.264 & 0.384 & 528.76 & 2500 \\
300 & TYP & QP & 1.333 & 0.000 & 0.000 & 0.000 & 0.564 & 11.328 & 35.03 & 152 \\
300 & TYP & RL & 15.920 & 0.000 & 0.005 & 0.731 & -4.811 & 0.545 & 6421.18 & 1420001 steps \\
\bottomrule\end{longtable}
\normalsize

\clearpage\subsection*{AC Cross-System and Scenarios}
\scriptsize
\begin{longtable}{lllrrrrrrrrrrrr}
\toprule
Case & Scenario & Method & MAE $P_g$  & MAE $V$  & Viol $P_g$ & Viol $P_g^0$ & Viol $Q_g$ & Viol $V$ & Viol line & Gap (\%) & PF (\%) & Infer (ms) & Train (s) & Epoch/steps \\
\midrule\endhead
30 & TYP & AS & 20.187 & 0.068 & 0.000 & 0.000 & 0.042 & 0.000 & 0.000 & 6.654 & 100.00 & 11.005 & 13.58 & 28 \\
30 & TYP & CP & 1.970 & 0.253 & 0.001 & 0.000 & 0.000 & 0.009 & 0.005 & -0.086 & 100.00 & 9.666 & 10240.01 & 272 \\
30 & TYP & DNN & 0.240 & 0.015 & 0.001 & 0.000 & 0.004 & 0.000 & 0.000 & 0.018 & 100.00 & 9.621 & 18.35 & 62 \\
30 & TYP & E-NGT & 32.507 & 0.105 & 0.104 & 0.000 & 0.004 & 0.000 & 0.007 & -5.565 & 100.00 & 10.903 & 6952.06 & 2500 \\
30 & TYP & FR & 0.000 & 0.001 & 0.000 & 0.000 & 0.000 & 0.000 & 0.000 & 0.000 & 100.00 & 232.519 & 10240.01 & 272 \\
30 & TYP & GNN & 3.245 & 3.576 & 0.002 & 0.000 & 0.490 & 0.000 & 0.061 & 0.514 & 100.00 & 22.332 & 85.57 & 66 \\
30 & TYP & KKT & 0.624 & 0.027 & 0.001 & 0.000 & 0.008 & 0.000 & 0.001 & -0.136 & 100.00 & 9.683 & 535.26 & 79 \\
30 & TYP & LR & 0.040 & 0.005 & 0.000 & 0.000 & 0.001 & 0.000 & 0.000 & 0.000 & 100.00 & 9.720 & 0.06 & -- \\
30 & TYP & MU & 0.569 & 0.032 & 0.001 & 0.000 & 0.002 & 0.001 & 0.002 & -0.121 & 100.00 & 9.973 & 199.88 & 80 \\
30 & TYP & NGT & 102.569 & 3.853 & 0.010 & 0.334 & 0.555 & 0.001 & 0.560 & -30.545 & 100.00 & 10.641 & 6308.90 & 2500 \\
30 & TYP & QC & 0.378 & 0.343 & 0.000 & 0.000 & 0.359 & 0.000 & 0.092 & 0.227 & 100.00 & 18.200 & 30.43 & 101 \\
30 & TYP & RL & 61.060 & 3.576 & 0.000 & 0.000 & 0.584 & 0.000 & 0.370 & -18.478 & 100.00 & 9.736 & 11561.49 & 800001 steps \\
118 & TYP & AS & 8.403 & 0.234 & 0.000 & 0.000 & 0.211 & 0.000 & 0.097 & 0.478 & 100.00 & 15.198 & 8.46 & 32 \\
118 & TYP & CP & 1.088 & 0.122 & 0.017 & 0.000 & 0.040 & 0.017 & 0.005 & -0.426 & 100.00 & 11.267 & 20795.64 & 479 \\
118 & TYP & DNN & 0.306 & 0.041 & 0.017 & 0.000 & 0.055 & 0.000 & 0.005 & -0.012 & 100.00 & 11.060 & 34.85 & 239 \\
118 & TYP & E-NGT & 15.509 & 0.396 & 0.118 & 0.012 & 0.081 & 0.000 & 0.272 & 1.500 & 100.00 & 12.489 & 4241.60 & 3000 \\
118 & TYP & FR & 0.000 & 0.001 & 0.000 & 0.000 & 0.000 & 0.000 & 0.000 & 0.000 & 100.00 & 563.989 & 20795.64 & 479 \\
118 & TYP & GNN & 0.958 & 3.702 & 0.004 & 0.000 & 1.218 & 0.000 & 0.072 & 0.403 & 100.00 & 33.259 & 134.97 & 127 \\
118 & TYP & KKT & 0.669 & 0.107 & 0.008 & 0.000 & 0.095 & 0.001 & 0.011 & -0.011 & 100.00 & 11.658 & 1153.03 & 330 \\
118 & TYP & LR & 0.518 & 0.066 & 0.025 & 0.000 & 0.066 & 0.001 & 0.002 & -0.002 & 100.00 & 17.921 & 4.55 & -- \\
118 & TYP & MU & 1.341 & 0.074 & 0.053 & 0.000 & 0.055 & 0.000 & 0.010 & 0.078 & 100.00 & 11.663 & 110.49 & 80 \\
118 & TYP & NGT & 13.976 & 3.328 & 0.089 & 0.010 & 0.177 & 0.000 & 0.228 & 1.482 & 100.00 & 12.413 & 3984.14 & 3000 \\
118 & TYP & QC & 0.272 & 0.065 & 0.000 & 0.000 & 0.068 & 0.000 & 0.004 & 0.000 & 100.00 & 21.702 & 38.07 & 186 \\
118 & TYP & RL & 14.718 & 3.702 & 0.000 & 0.000 & 1.217 & 0.000 & 0.109 & 3.817 & 100.00 & 11.183 & 9804.70 & 570001 steps \\
118 & API & AS & 46.913 & 4.014 & 0.000 & 19.557 & 0.439 & 0.000 & 1.884 & -28.356 & 100.00 & 15.372 & 8.46 & 32 \\
118 & API & CP & 69.080 & 3.794 & 2.060 & 35.631 & 1.598 & 0.013 & 2.571 & -24.733 & 100.00 & 13.448 & 20795.64 & 479 \\
118 & API & DNN & 87.063 & 4.449 & 7.379 & 17.248 & 1.380 & 0.058 & 1.244 & -25.546 & 100.00 & 13.162 & 34.85 & 239 \\
118 & API & E-NGT & 77.568 & 4.036 & 3.761 & 24.946 & 0.438 & 0.007 & 1.578 & -28.562 & 100.00 & 14.401 & 4241.60 & 3000 \\
118 & API & FR & 0.000 & 0.000 & 0.000 & 0.000 & 0.000 & 0.000 & 0.000 & 0.000 & 100.00 & 600.395 & 20795.64 & 479 \\
118 & API & GNN & NR & NR & NR & NR & NR & NR & NR & NR & 0.00 & 46.697 & 134.97 & 127 \\
118 & API & KKT & 90.142 & 5.823 & 7.453 & 10.208 & 1.972 & 0.066 & 1.673 & -28.766 & 100.00 & 13.722 & 1153.03 & 330 \\
118 & API & LR & 100.550 & 8.404 & 16.547 & 11.815 & 6.676 & 0.212 & 3.375 & -20.298 & 100.00 & 19.880 & 4.55 & -- \\
118 & API & MU & 77.380 & 4.108 & 2.647 & 24.977 & 0.481 & 0.032 & 1.682 & -27.626 & 100.00 & 13.759 & 110.49 & 80 \\
118 & API & NGT & 76.496 & 3.832 & 3.204 & 22.965 & 0.196 & 0.000 & 1.397 & -31.051 & 100.00 & 12.815 & 3984.14 & 3000 \\
118 & API & QC & 77.627 & 3.781 & 6.600 & 27.588 & 0.059 & 0.018 & 1.806 & -24.696 & 100.00 & 25.486 & 38.07 & 186 \\
118 & API & RL & 47.918 & 3.562 & 0.000 & 17.106 & 1.243 & 0.006 & 1.999 & -26.566 & 100.00 & 11.330 & 9804.70 & 570001 steps \\
118 &  VAR & AS & 8.227 & 0.444 & 0.000 & 0.004 & 0.245 & 0.000 & 0.157 & 0.246 & 100.00 & 15.431 & 8.46 & 32 \\
118 &  VAR & CP & 1.919 & 0.212 & 0.015 & 0.000 & 0.131 & 0.013 & 0.032 & -0.352 & 100.00 & 11.123 & 20795.64 & 479 \\
118 &  VAR & DNN & 1.549 & 0.161 & 0.036 & 0.000 & 0.181 & 0.002 & 0.033 & -0.078 & 100.00 & 11.198 & 34.85 & 239 \\
118 &  VAR & E-NGT & 19.849 & 0.538 & 0.268 & 0.121 & 0.140 & 0.000 & 0.282 & 2.438 & 100.00 & 12.399 & 4241.60 & 3000 \\
118 &  VAR & FR & 0.000 & 0.001 & 0.000 & 0.000 & 0.000 & 0.000 & 0.000 & 0.000 & 100.00 & 556.321 & 20795.64 & 479 \\
118 &  VAR & GNN & 7.153 & 3.731 & 0.811 & 0.008 & 1.224 & 0.000 & 0.117 & -1.781 & 99.99 & 32.648 & 134.97 & 127 \\
118 &  VAR & KKT & 2.267 & 0.298 & 0.066 & 0.000 & 0.295 & 0.005 & 0.034 & -0.146 & 100.00 & 11.535 & 1153.03 & 330 \\
118 &  VAR & LR & 1.917 & 0.202 & 0.249 & 0.000 & 0.252 & 0.004 & 0.029 & -0.063 & 100.00 & 17.782 & 4.55 & -- \\
118 &  VAR & MU & 2.932 & 0.198 & 0.077 & 0.000 & 0.142 & 0.001 & 0.038 & 0.028 & 100.00 & 11.719 & 110.49 & 80 \\
118 &  VAR & NGT & 19.177 & 3.352 & 0.259 & 0.096 & 0.213 & 0.000 & 0.234 & 2.307 & 100.00 & 12.443 & 3984.14 & 3000 \\
118 &  VAR & QC & 0.980 & 0.159 & 0.000 & 0.000 & 0.175 & 0.001 & 0.023 & -0.021 & 100.00 & 21.742 & 38.07 & 186 \\
118 &  VAR & RL & 15.179 & 3.731 & 0.000 & 0.001 & 1.215 & 0.000 & 0.128 & 3.714 & 100.00 & 11.221 & 9804.70 & 570001 steps \\
300 & TYP & AS & 26.604 & 0.442 & 0.000 & 0.217 & 14.666 & 0.201 & 0.989 & -5.625 & 3.30 & 36.817 & 3.69 & 26 \\
300 & TYP & CP & 10.718 & 0.797 & 0.129 & 0.003 & 0.374 & 0.074 & 0.106 & 1.011 & 100.00 & 17.868 & 8973.62 & 130 \\
300 & TYP & DNN & 3.995 & 0.216 & 0.216 & 0.002 & 0.217 & 0.004 & 0.013 & 0.701 & 100.00 & 17.967 & 19.73 & 246 \\
300 & TYP & E-NGT & 15.155 & 0.475 & 1.850 & 0.423 & 0.528 & 0.010 & 0.089 & 1.152 & 100.00 & 19.283 & 2033.66 & 2500 \\
300 & TYP & FR & 1.175 & 0.088 & 0.014 & 0.000 & 0.041 & 0.008 & 0.011 & 0.117 & 100.00 & 1553.091 & 8973.62 & 130 \\
300 & TYP & GNN & 6.566 & 3.924 & 0.378 & 0.474 & 3.786 & 0.061 & 0.093 & 0.062 & 37.80 & 68.736 & 318.52 & 212 \\
300 & TYP & KKT & 4.150 & 0.522 & 0.222 & 0.000 & 0.889 & 0.008 & 0.008 & 0.033 & 100.00 & 18.275 & 478.25 & 273 \\
300 & TYP & LR & 4.565 & 0.248 & 0.409 & 0.000 & 0.145 & 0.006 & 0.011 & -0.028 & 100.00 & 26.831 & 15.25 & -- \\
300 & TYP & MU & 8.067 & 0.595 & 0.034 & 1.383 & 1.046 & 0.025 & 0.204 & 0.029 & 98.50 & 18.957 & 74.70 & 80 \\
300 & TYP & NGT & 102.907 & 4.366 & 13.925 & 0.134 & 3.625 & 0.063 & 0.173 & 5.855 & 100.00 & 19.322 & 1951.31 & 2500 \\
300 & TYP & QC & 3.205 & 0.291 & 0.000 & 0.017 & 0.232 & 0.014 & 0.042 & 0.291 & 99.30 & 36.565 & 32.78 & 256 \\
300 & TYP & RL & 27.523 & 3.919 & 0.000 & 0.000 & 3.528 & 0.067 & 0.517 & 4.725 & 100.00 & 17.799 & 20500.75 & 780001 steps \\
\bottomrule\end{longtable}
\normalsize

\clearpage\subsection*{DC Data Scaling}
\scriptsize
\begin{longtable}{lllrrrrrrrr}
\toprule
Size & Scenario & Method & MAE $P_g$  & Viol $P_g$ & Viol $P_g^0$ & Viol line & Gap (\%) & Infer (ms) & Train (s) & Epoch/steps \\
\midrule\endhead
1000 & TYP & DNN & 1.790 & 0.001 & 0.000 & 0.043 & -0.002 & 0.437 & 0.86 & 48 \\
1000 & TYP & KKT & 1.378 & 0.022 & 0.000 & 0.039 & -0.035 & 0.456 & 15.36 & 218 \\
1000 & TYP & LR & 1.008 & 0.074 & 0.000 & 0.004 & -0.003 & 0.118 & 0.01 & -- \\
1000 & TYP & MU & 1.740 & 0.000 & 0.000 & 0.033 & 0.039 & 0.458 & 2.62 & 80 \\
1000 & TYP & QP & 2.187 & 0.000 & 0.000 & 0.000 & 0.566 & 5.042 & 1.79 & 61 \\
5000 & TYP & DNN & 1.014 & 0.035 & 0.000 & 0.014 & 0.007 & 0.442 & 8.68 & 123 \\
5000 & TYP & KKT & 1.321 & 0.011 & 0.000 & 0.035 & -0.027 & 0.452 & 25.80 & 75 \\
5000 & TYP & LR & 0.979 & 0.065 & 0.000 & 0.004 & 0.002 & 0.121 & 0.04 & -- \\
5000 & TYP & MU & 0.911 & 0.000 & 0.000 & 0.011 & 0.039 & 0.460 & 11.78 & 80 \\
5000 & TYP & QP & 0.891 & 0.000 & 0.000 & 0.000 & 0.219 & 4.774 & 15.15 & 124 \\
12000 & TYP & DNN & 0.725 & 0.024 & 0.000 & 0.017 & 0.002 & 0.438 & 14.90 & 92 \\
12000 & TYP & KKT & 0.979 & 0.014 & 0.000 & 0.022 & -0.033 & 0.449 & 55.51 & 68 \\
12000 & TYP & LR & 0.970 & 0.065 & 0.000 & 0.004 & 0.001 & 0.121 & 0.09 & -- \\
12000 & TYP & MU & 0.330 & 0.000 & 0.000 & 0.004 & 0.015 & 0.471 & 27.83 & 80 \\
12000 & TYP & QP & 0.413 & 0.000 & 0.000 & 0.000 & 0.122 & 5.330 & 29.45 & 103 \\
20000 & TYP & DNN & 0.457 & 0.025 & 0.000 & 0.004 & 0.012 & 0.437 & 28.30 & 103 \\
20000 & TYP & KKT & 0.139 & 0.006 & 0.000 & 0.002 & 0.000 & 0.443 & 247.03 & 182 \\
20000 & TYP & LR & 0.968 & 0.066 & 0.000 & 0.004 & 0.000 & 0.121 & 0.15 & -- \\
20000 & TYP & MU & 0.281 & 0.000 & 0.000 & 0.003 & 0.018 & 0.468 & 45.95 & 80 \\
20000 & TYP & QP & 0.371 & 0.000 & 0.000 & 0.000 & 0.096 & 5.227 & 39.80 & 83 \\
30000 & TYP & DNN & 0.293 & 0.011 & 0.000 & 0.004 & -0.001 & 0.440 & 44.65 & 109 \\
30000 & TYP & KKT & 0.149 & 0.006 & 0.000 & 0.003 & -0.008 & 0.445 & 678.88 & 332 \\
30000 & TYP & LR & 0.968 & 0.066 & 0.000 & 0.004 & 0.000 & 0.119 & 0.22 & -- \\
30000 & TYP & MU & 0.223 & 0.000 & 0.000 & 0.008 & 0.002 & 0.474 & 67.97 & 80 \\
30000 & TYP & QP & 0.240 & 0.000 & 0.000 & 0.000 & 0.040 & 4.544 & 66.03 & 94 \\
35000 & TYP & DNN & 0.324 & 0.013 & 0.000 & 0.004 & 0.004 & 0.448 & 39.92 & 83 \\
35000 & TYP & KKT & 0.253 & 0.018 & 0.000 & 0.001 & 0.001 & 0.447 & 316.21 & 130 \\
35000 & TYP & LR & 0.968 & 0.066 & 0.000 & 0.004 & 0.000 & 0.118 & 0.27 & -- \\
35000 & TYP & MU & 0.242 & 0.000 & 0.000 & 0.005 & -0.004 & 0.468 & 74.20 & 80 \\
35000 & TYP & QP & 0.266 & 0.000 & 0.000 & 0.000 & 0.081 & 5.414 & 93.96 & 117 \\
\bottomrule\end{longtable}
\normalsize

\subsection*{AC Data Scaling}
\scriptsize
\begin{longtable}{lllrrrrrrrrrrrr}
\toprule
Size & Scenario & Method & MAE $P_g$ & MAE $V$ & Viol $P_g$ & Viol $P_g^0$ & Viol $Q_g$ & Viol $V$ & Viol line & Gap (\%) & PF (\%) & Infer (ms) & Train (s) & Epoch/steps \\
\midrule\endhead
1000 & TYP & DNN & 1.286 & 0.120 & 0.018 & 0.000 & 0.098 & 0.002 & 0.013 & 0.018 & 100.00 & 11.178 & 5.67 & 363 \\
1000 & TYP & KKT & 2.443 & 0.205 & 0.039 & 0.000 & 0.194 & 0.002 & 0.027 & 0.011 & 100.00 & 11.399 & 112.63 & 309 \\
1000 & TYP & LR & 0.615 & 0.081 & 0.031 & 0.000 & 0.096 & 0.001 & 0.003 & -0.002 & 100.00 & 17.832 & 0.82 & -- \\
1000 & TYP & MU & 2.082 & 0.147 & 0.013 & 0.000 & 0.086 & 0.002 & 0.034 & 0.006 & 100.00 & 11.843 & 22.69 & 80 \\
1000 & TYP & QC & 2.988 & 0.233 & 0.000 & 0.000 & 0.029 & 0.000 & 0.049 & 0.141 & 100.00 & 21.419 & 2.62 & 104 \\
5000 & TYP & DNN & 0.535 & 0.056 & 0.017 & 0.000 & 0.074 & 0.001 & 0.006 & -0.002 & 100.00 & 11.228 & 13.08 & 175 \\
5000 & TYP & KKT & 1.404 & 0.145 & 0.022 & 0.000 & 0.120 & 0.001 & 0.006 & -0.048 & 100.00 & 11.609 & 425.84 & 247 \\
5000 & TYP & LR & 0.529 & 0.070 & 0.026 & 0.000 & 0.075 & 0.001 & 0.002 & -0.002 & 100.00 & 17.882 & 2.54 & -- \\
5000 & TYP & MU & 2.336 & 0.103 & 0.058 & 0.000 & 0.070 & 0.000 & 0.013 & 0.118 & 100.00 & 11.816 & 83.04 & 80 \\
5000 & TYP & QC & 0.302 & 0.065 & 0.000 & 0.000 & 0.086 & 0.000 & 0.003 & 0.014 & 100.00 & 21.580 & 61.64 & 334 \\
12000 & TYP & DNN & 0.287 & 0.039 & 0.011 & 0.000 & 0.054 & 0.000 & 0.003 & 0.008 & 100.00 & 11.160 & 41.75 & 231 \\
12000 & TYP & KKT & 0.594 & 0.110 & 0.010 & 0.000 & 0.087 & 0.001 & 0.008 & -0.036 & 100.00 & 11.487 & 1754.40 & 437 \\
12000 & TYP & LR & 0.515 & 0.068 & 0.026 & 0.000 & 0.072 & 0.001 & 0.002 & -0.003 & 100.00 & 17.905 & 6.06 & -- \\
12000 & TYP & MU & 1.163 & 0.069 & 0.038 & 0.000 & 0.049 & 0.001 & 0.017 & 0.081 & 100.00 & 11.830 & 153.44 & 80 \\
12000 & TYP & QC & 0.221 & 0.044 & 0.000 & 0.000 & 0.087 & 0.000 & 0.001 & 0.013 & 100.00 & 21.356 & 95.84 & 373 \\
20000 & TYP & DNN & 0.217 & 0.038 & 0.006 & 0.000 & 0.087 & 0.000 & 0.002 & 0.002 & 100.00 & 11.208 & 60.49 & 209 \\
20000 & TYP & KKT & 0.479 & 0.101 & 0.010 & 0.000 & 0.092 & 0.001 & 0.005 & 0.001 & 100.00 & 11.560 & 2419.66 & 363 \\
20000 & TYP & LR & 0.514 & 0.068 & 0.025 & 0.000 & 0.071 & 0.001 & 0.002 & -0.003 & 100.00 & 17.878 & 9.86 & -- \\
20000 & TYP & MU & 0.568 & 0.051 & 0.007 & 0.000 & 0.033 & 0.000 & 0.006 & 0.063 & 100.00 & 11.722 & 267.81 & 80 \\
20000 & TYP & QC & 0.191 & 0.040 & 0.000 & 0.000 & 0.070 & 0.000 & 0.003 & 0.003 & 100.00 & 21.601 & 148.23 & 270 \\
30000 & TYP & DNN & 0.173 & 0.025 & 0.005 & 0.000 & 0.048 & 0.000 & 0.004 & -0.009 & 100.00 & 11.261 & 127.26 & 289 \\
30000 & TYP & KKT & 0.403 & 0.089 & 0.006 & 0.000 & 0.099 & 0.001 & 0.006 & -0.036 & 100.00 & 11.600 & 4356.19 & 420 \\
30000 & TYP & LR & 0.515 & 0.068 & 0.025 & 0.000 & 0.071 & 0.001 & 0.002 & -0.002 & 100.00 & 17.776 & 15.08 & -- \\
30000 & TYP & MU & 0.462 & 0.043 & 0.006 & 0.000 & 0.025 & 0.000 & 0.005 & 0.058 & 100.00 & 11.809 & 353.50 & 80 \\
30000 & TYP & QC & 0.163 & 0.038 & 0.000 & 0.000 & 0.070 & 0.000 & 0.002 & 0.002 & 100.00 & 21.650 & 102.69 & 226 \\
35000 & TYP & DNN & 0.216 & 0.031 & 0.007 & 0.000 & 0.056 & 0.001 & 0.002 & -0.018 & 100.00 & 11.337 & 111.59 & 217 \\
35000 & TYP & KKT & 0.306 & 0.075 & 0.004 & 0.000 & 0.075 & 0.001 & 0.006 & -0.001 & 100.00 & 11.667 & 6052.03 & 519 \\
35000 & TYP & LR & 0.515 & 0.068 & 0.025 & 0.000 & 0.071 & 0.001 & 0.002 & -0.002 & 100.00 & 18.180 & 18.03 & -- \\
35000 & TYP & MU & 0.472 & 0.039 & 0.004 & 0.000 & 0.027 & 0.000 & 0.005 & 0.050 & 100.00 & 11.638 & 444.71 & 80 \\
35000 & TYP & QC & 0.169 & 0.037 & 0.000 & 0.000 & 0.086 & 0.000 & 0.002 & 0.008 & 100.00 & 21.768 & 172.77 & 202 \\
\bottomrule\end{longtable}
\normalsize

\clearpage\subsection*{Conventional Solver Reference Timing}
\begin{longtable}{llr}\toprule
Formulation & Case & Solver time (ms) \\
\midrule\endhead
DC & 30 & 14.88 \\
DC & 118 & 419.91 \\
DC & 300 & 3325.96 \\
AC & 30 & 67.32 \\
AC & 118 & 432.57 \\
AC & 300 & 14093.14 \\
\bottomrule\end{longtable}

\section{Direct Network Outputs}
\label{app:direct}

The main benchmark uses the non-slack active-generation
and generator-voltage setpoints predicted by \texttt{MU}
and \texttt{KKT}, and obtains the remaining variables
through AC power flow.
This appendix compares their direct network outputs
with the resulting power-flow states to examine how
reconstruction affects prediction accuracy.
Both evaluations use the same frozen networks,
without retraining.

In the Direct-Output and Power-Flow Comparison Table, MAE$_{P_g}$ covers non-slack generators, MAE$_{V_m}$ covers
generator buses, and MAE$_{Q_g}$ covers all generators.
These three errors use the percentage normalization adopted
in the main benchmark. A single column is reported for
MAE$_{P_g}$ and MAE$_{V_m}$ because both stages use the same
predicted setpoints.
Voltage-angle error MAE$_{\theta}$ is averaged over all buses
and matched samples, after aligning predicted and reference
angles to their respective reference-bus angles and wrapping
their differences to $[-\pi,\pi]$. It is reported in degrees.

Direct reactive-generation predictions have lower errors than
the power-flow results for both methods on all three systems.
Power flow reduces angle error on the 30-bus system but
increases it on the 118- and 300-bus systems.
The power-flow stage recomputes reactive generation and the
remaining voltage state to satisfy the network equations
at the predicted setpoints; it does not minimize their
distance from the reference OPF solution.
Consequently, restoring power balance can increase prediction
error, and power-flow convergence alone does not guarantee
that all operating limits are satisfied.

\subsection*{Direct-Output and Power-Flow Comparison}
\scriptsize
\begin{table}[H]
\centering
\label{tab:direct-pf-comparison}
\scriptsize
\setlength{\tabcolsep}{3pt}
\resizebox{\linewidth}{!}{%
\begin{tabular}{llrrrrrrr}
\toprule
System & Method & Samples
& \shortstack{MAE$_{P_g}$\\(\%)}
& \shortstack{MAE$_{V_m}$\\(\%)}
& \shortstack{Direct MAE$_{Q_g}$\\(\%)}
& \shortstack{PF MAE$_{Q_g}$\\(\%)}
& \shortstack{Direct MAE$_{\theta}$\\($^\circ$)}
& \shortstack{PF MAE$_{\theta}$\\($^\circ$)} \\
\midrule
30-bus  & MU  & 1000 & 0.569 & 0.032 & 0.407 & 2.550
        & 0.029 & 0.016 \\
        & KKT & 1000 & 0.624 & 0.027 & 0.587 & 2.536
        & 0.035 & 0.007 \\
\midrule
118-bus & MU  & 1000 & 1.341 & 0.074 & 3.620 & 5.058
        & 0.241 & 0.269 \\
        & KKT & 1000 & 0.669 & 0.107 & 3.100 & 8.359
        & 0.185 & 0.254 \\
\midrule
300-bus & MU  & 985 & 8.067 & 0.595 & 7.665 & 13.664
        & 12.282 & 15.973 \\
        & KKT & 1000 & 4.150 & 0.522 & 4.383 & 10.563
        & 2.849 & 4.925 \\
\bottomrule
\end{tabular}%
}
\end{table}

\end{document}